\documentclass{jfm}

\usepackage[latin1]{inputenc}
\usepackage[english]{babel}
\usepackage{float}
\usepackage{subfig}
\usepackage{textcomp}
\usepackage{amsmath}
\usepackage{graphicx}
\usepackage{setspace}
\usepackage{esint}
\usepackage{natbib}
\usepackage{color,soul}

\usepackage{psfrag}

\usepackage[T1]{fontenc}               
\usepackage[section]{placeins}         

\usepackage{fancyhdr}
\usepackage{mathrsfs}
\usepackage{amsfonts}
\usepackage{lscape}
\usepackage{longtable}

\usepackage{array}
\newcolumntype{L}[1]{>{\raggedright\let\newline\\\arraybackslash\hspace{0pt}}m{#1}}
\newcolumntype{C}[1]{>{\centering\let\newline\\\arraybackslash\hspace{0pt}}m{#1}}
\newcolumntype{R}[1]{>{\raggedleft\let\newline\\\arraybackslash\hspace{0pt}}m{#1}}

\ifCUPmtlplainloaded \else
  \checkfont{eurm10}
  \iffontfound
    \IfFileExists{upmath.sty}
      {\typeout{^^JFound AMS Euler Roman fonts on the system,
                   using the 'upmath' package.^^J}%
       \usepackage{upmath}}
      {\typeout{^^JFound AMS Euler Roman fonts on the system, but you
                   dont seem to have the}%
       \typeout{'upmath' package installed. JFM.cls can take advantage
                 of these fonts,^^Jif you use 'upmath' package.^^J}%
      }
  \else
  \fi
\fi

\ifCUPmtlplainloaded \else
  \checkfont{msam10}
  \iffontfound
    \IfFileExists{amssymb.sty}
      {\typeout{^^JFound AMS Symbol fonts on the system, using the
                'amssymb' package.^^J}%
       \usepackage{amssymb}%

      }{}
  \fi
\fi

\ifCUPmtlplainloaded \else
  \IfFileExists{amsbsy.sty}
    {\typeout{^^JFound the 'amsbsy' package on the system, using it.^^J}%
     \usepackage{amsbsy}}
    {}
\fi

\newsavebox{\astrutbox}
\sbox{\astrutbox}{\rule[-5pt]{0pt}{20pt}}

\title[Crest speeds of unsteady water waves]{Crest speeds of unsteady surface water waves}

\author[F. Fedele, M. L. Banner, X. Barthelemy] 
{Francesco Fedele$^1$ \thanks{Email address for correspondence: fedele@gatech.edu},
 Michael L. Banner$^2$, Xavier Barthelemy$^{2,3}$}   

\affiliation{$^1$School of Civil and Environmental Engineering, Georgia Institute of Technology, Atlanta, GA 30332, USA\\$^2$School of Mathematics and Statistics, UNSW Australia, Sydney NSW 2052, Australia\\[\affilskip]
$^3$Water Research Laboratory, School of Civil and Environmental Engineering, UNSW Australia, Sydney NSW 2052, Australia.}

\pubyear{2018}
\volume{xxx}
\pagerange{xxx--xxx}
\date{?; revised ?; accepted ?. - To be entered by editorial office}

\begin{document}
\maketitle 
\begin{abstract}
Intuitively, crest speeds of water waves are assumed to match their phase speeds. However, this is generally not the case for natural waves within unsteady wave groups. This motivates our study, which presents new insights into the generic behavior of crest speeds of linear to highly nonlinear unsteady waves. While our major focus is on gravity waves where a generic crest slowdown occurs cyclically, results for capillary-dominated waves are also discussed, for which crests cyclically speed up. This curious phenomenon arises when the theoretical constraint of steadiness is relaxed, allowing waves to change their form, or shape. In particular, a kinematic analysis of both simulated and observed open ocean gravity waves reveals a forward-to-backward leaning cycle for each individual crest within a wave group. This is clearly manifest during the focusing of dominant wave groups essentially due to the dispersive nature of waves. It occurs routinely for focusing linear (vanishingly small steepness) wave groups, and it is enhanced as the wave spectrum broadens. It is found to be relatively insensitive to the degree of phase coherence and focusing of wave groups. The nonlinear nature of waves limits the crest slowdown. This reduces when gravity waves become less dispersive, either as they steepen or as they propagate over finite water depths.  This is demonstrated by numerical simulations of the unsteady evolution of 2D and 3D dispersive gravity wave packets in both deep and intermediate water depths, and by open ocean space-time measurements.
\end{abstract}

\section{Introduction}

Wind-generated ocean waves at the air-sea interface propagate predominantly in groups. 
Unsteady wave groups can exhibit a complex nonlinear life cycle, especially in focal zones where there is a rapid concentration of wave energy. Their structure and the associated carrier wave propagation depend on the dispersion, directionality and nonlinearity of waves, and their interplay. Understanding natural ocean wave propagation is not only of academic relevance, but it is also of importance in studying the energy exchange between the wind and waves~\citep{Sullivan2018,LiberzonShemer2011}. Crest speeds are a key consideration for assessing the onset of breaking of a wave. In particular, \citet{Barthelemy2018} identified a new breaking onset threshold criterion based on energy flux considerations, which is valid for 2D and 3D wave groups in deep and intermediate water depths. While the energy flux that initiates breaking arises from energy focusing within the whole wave group, they found that the initial instability, or point of no return beyond which waves inevitably break, occurs within a very compact region centred on the maximum wave crest. Their new breaking criterion is based on the strength of the local energy flux relative to the local energy density, normalised by the local crest speed $c$. On the free surface, at the crest point, their dynamical criterion reduces to $B_x=u/c$, a kinematic condition for the ratio of crest fluid speed $u$ to crest point speed $c$, where $x$ is the direction of propagation. For maximally steep non-breaking waves, the crest fluid speed $u$ cannot exceed $0.85$ of the crest speed $c$, i.e. $B_x=0.85$.  In any event, their breaking onset threshold occurs well before the water speed outruns the crest speed.

In other applications, measurements of whitecap speeds have been proposed as a method for inferring ocean wavelengths~\citep{Phillips1985}, assuming that the Stokes dispersion relationship is applicable. Any systematic crest speed slowdown just prior to breaking onset can significantly alias wavelengths determined using the quadratic dispersion relation between Stokes wavelength and wave speed. Moreover, in the design of fixed and floating offshore structures, whitecap speeds are a key consideration in assessing wave impact loading on such structures, where flow-induced forces are proportional to the square of the water velocity. In particular,  an $O(20\%)$ crest slowdown of breaking crests will result in a $O( 50\%)$ reduction in their momentum flux loading. For nonbreaking waves, crest slowdown effects will likely be much weaker.


\cite{Stokes1847} classical deep-water wave theory was developed for a steady, uniform train of two-dimensional (2D) nonlinear, deep-water waves of small-to-intermediate mean steepness $a k\left(=2\pi a / L \right)$, where $a$ is the wave amplitude and $L$ is the wavelength. The phase speed $c$ increases with $a k$ as found to $5^{th}$~order in wave steepness~(see, e.g.,~\cite{Fenton1985}):

\begin{equation}
c=c_{0}\left(1+\frac{1}{2}\left(ak\right)^{2}+\frac{1}{8}\left(ak\right)^{4}+O((ak)^6)\right)\label{eq:Stokes}
\end{equation}
where $c_{0}$ is the wave speed of linear (infinitesimally steep) waves.

The limiting phase speed of $1.1c_{0}$ for maximally-steep steady waves follows by increasing the steepness up to the Stokes limit ($a k=\pi/7\approx0.449$). Increased wave steepness has long been associated with higher wave speeds. Historically, Stokes theory has been the default theoretical approach for describing ocean waves~\citep{Kinsman1965,Wiegel1964}. Contemporary studies by~\citet{Beya2012} found close agreement between monochromatic wave near-surface velocities and predictions based on $5^{\mbox{th}}$~order Stokes theory. Recent observational studies of velocities under extreme and breaking crests showed that inviscid predictions are in good agreement with PIV-measured particle velocities~\citep{Grue2003}. Further, \citet{Johannessen2010} showed that at the crest of a nonbreaking wave, near-surface velocities agree with second-order Stokes theory, which is capable of describing the free surface kinematics very accurately provided that the local underlying regime of free waves can be identified. A principal constraint of the Stokes wavetrain framework is the imposed uniform spatial and temporal periodicity, which provides the basis for the extensive analytic treatment in the literature of this intrinsically nonlinear free surface problem.

However, ocean waves propagate naturally in unsteady groups that evolve dynamically. Relaxing the periodicity constraints allows the manifestation of additional degrees of freedom, as reported in a number of laboratory and numerical wave tank studies. For example, \citet{Melville1983} found a phase speed variation of between $-17\%$ and $+32\%$ around the linear phase velocity. \citet{Shemer2013} demonstrated that the crest propagation speed of dominant waves in broadbanded groups differs significantly from both their phase and group velocities. More specifically, crest slowdown phenomena have been observed for gravity waves and studied as a phase-resolved wave effect by several authors~\citep{Johannessen2001, Johannessen2003, Johannessen2010, Katsardi2011, Shemer2013, Shemer2014, Shemer2015}. Deep-water breaking wave studies all found a significant~$\sim\mathcal{O}(20\%)$ breaking crest speed slowdown relative to the expected linear phase velocity~\citep{Rapp1990, Stansell2002, Jessup2005, Melville2002, MELVILLE2002a}.

Understanding gravity wave crest slowdown behavior is central to the refining of the present knowledge of water-wave propagation and dynamics, and also to an effective implementation of Phillips' spectral framework for breaking waves~\citep{Phillips1985,Gemmrich2008,Kleiss2010a}. The groupiness of natural ocean gravity waves gives rise to a suite of differences from Stokes wave predictions. \citet{Baldock1996} report results of a laboratory study of an unsteady packet produced by an ensemble of unidirectional wave modes of different frequency components focusing in space and time. Subsequently, \citet{Johannessen2001} extended that study to directional waves. Surface elevations and subsurface particle kinematics were compared with linear wave theory and the~\citet{Longuet-Higgins1960a} second-order solution of the wave-wave interactions. The study in~\citet{Baldock1996} shows that nonlinear wave-wave interactions produce a highly nonlinear wave group with crests higher and troughs shallower than expected from numerical simulations by~\citet{Longuet-Higgins1987} and experiments by~\citet{Miller1991}. \citet{Sutherland1995} previously obtained very similar characterisations, although they also concluded that the wave kinematics were Stokes-like. \citet{Song2002},~\citet{Banner2007} and~\citet{Viotti2014} report complementary numerical and experimental studies demonstrating that intra-group and inter-wave energy transfers cannot be neglected when characterizing wave group kinematics and dynamics.

Clearly,  a Stokes characterization of waves has significant limitations when applied to fully-nonlinear unsteady wave groups. Specifically, important properties derived from~\cite{Stokes1847}~theory and subsequent refinements by~\cite{Fenton1985} change fundamentally when stationarity is relaxed. Under gravity wave crest maxima, the velocity profiles differ systematically from those of a comparably steep Stokes wave. The unsteady leaning of waves~\citep{Tayfun1986} is shown to be a critical aspect, determining the actual wave crest speed, now unsteady. In~\citet{Banner2014}, our preliminary numerical study of chirped wave groups provides an overview of the gravity wave crest slowdown phenomenon, and have already been verified against observations of wave groups in the laboratory and in the open ocean field~(see also~\cite{Fedele2014}). This generic crest slowdown phenomenon has been recently observed by other researchers~\citep{Shemer2014,Liberzon2017,Shemer2018,Craciunescu2019,Craciunescu2019a}. It is investigated in much greater detail in the present paper.

\begin{figure}
\centering\includegraphics[scale=0.55]{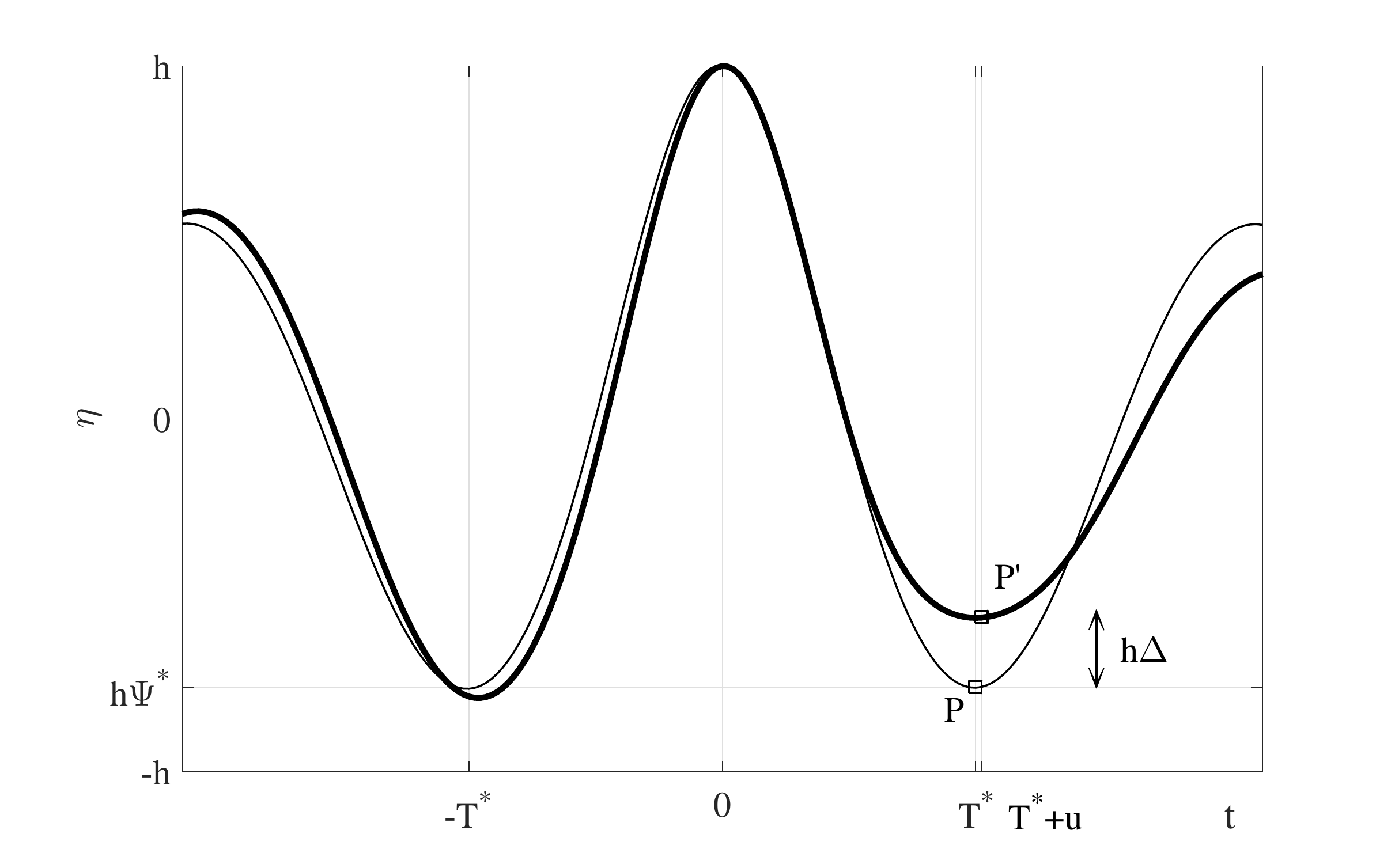} 
\caption{Time profile $\zeta(x=0,t)$ of the stochastic wave group of Eq.~\eqref{etac} at the focusing point ($x=0$). The thin line is the scaled covariance function $h\psi(t)$, which represents the symmetric wave profile of a crest in the limit of infinite amplitude $h/\sigma\rightarrow\infty$. The crest is followed by a trough at $T=T^*$ (point P) and amplitude $\eta_{tr}=h\psi^*$. The preceding trough occurs at $t=-T^*$ because of symmetry of $\psi(t)$. For finite but large crest amplitudes ($h/\sigma\gg1$), the wave profile is depicted by the bold line. The trough now occurs at $t=T^*+u$ (point P') with $u$ of order~$O(h^{-1})$ (negligible to $O(h^0)$, see~\citep{FEDELE2009}) and the amplitude $\eta_{tr}=h(\psi^*+\Delta)$, with $h\Delta$ of~$O(h^0)$~\citep{FEDELE2009}. Note that for $ \Delta\neq0$ the symmetry of the wave profile (bold line) is broken.}
\label{FIG1}
\end{figure}

\section{Crest slowdown of linear gravity wave groups}

Intuitively, crest speeds of water waves are assumed to match their phase speeds. However, our recent study highlights that this is generally not the case for natural water waves, which mostly occur in unsteady groups~\citep{Banner2014}.  Wave crest slowdown is observed as the manifestation of the natural dispersion of waves~\citep{Fedele2014}. It arises when the theoretical constraint of steadiness is relaxed, allowing  additional asymmetric degrees of freedom in the wave motion. Generically, at focusing sites longer deep-water wave components overtaking shorter wave components reinforce to create a locally large wave. The crest of this wave undergoes a slowdown as it approaches its maximum amplitude, followed by a speed-up during the subsquent decaying phase. We note that in strong contrast, in capillary wave packets, Fourier components with higher wave numbers (shorter wavelengths) have larger phase speeds and overtake the slower-propagating longer capillary wave components. In this case, the crest speeds up as it attains its maximum amplitude, and then slows down. The wave crest height at focus depends on the packet amplitude spectrum (bandwidth and shape) and the degree of phase coherence during focusing. This occurs at a point when the tallest wave in group attains its maximum possible crest height by a constructive superposition of elementary harmonic waves, whose amplitudes depend upon the assumed spectrum. The full or partial coherence of their phases decides whether the focusing is perfect or imperfect, but the crest slowdown is practically insensitive to the degree of focusing, as discussed below. In the following, our study concentrates predominantly on gravity waves, but we will also consider waves in the capillary range for completeness.

\subsection{Perfect focusing}


Consider generic dispersive waves of an unidirectional Gaussian sea with frequency spectrum $E(\omega)$ and variance $\sigma^2$. The associated wavenumber spectrum $S(k)=E(\omega)\frac{d\omega}{dk}$, where $\omega(k)$ is the linear dispersion relation. Later, we will particularize our results to both deep and finite depth water waves as well as capillary waves. A measure of the spectral bandwidth of the Gaussian sea can be defined as $\nu=\sqrt{m_0 m_2/m_1^2-1}$, where the spectral moments $m_{j}=\int \omega^{j}E(\omega)d\omega$. The dominant wavenumber is $k_{0}$ and the associated frequency $\omega_{0}=\omega(k_0)$.  

Assume that a large crest of amplitude $h$ is observed at $x = 0$ and $t = 0$. \cite{Boccotti2000}~showed that as $h/\sigma\rightarrow\infty$, the large crest belongs to a well-defined wave group that passes through $x=0$ and it reaches its maximum focus at time $t=0$. The surface displacement $\zeta$ of the group around focus is described, correct to $O(h^0)$, by the conditional random process~\citep{Fedele2008,FEDELE2009}
\begin{equation}
\zeta(x,t)=\left\{\eta(x,t)\vert\eta(0,0)=h\right\} =h \left\{\Psi(x,t)+\Delta\frac{\Psi(x,t-T^*)-\psi^{*}\Psi(x,t)}{1-\psi^{*2}}\right\},\label{etac}
\end{equation}
where
\begin{equation}
\Psi(x,t)=\frac{\overline{\eta(\xi,\tau)\eta(\xi+x,\tau+t)}}{\sigma^2}=\int_0^{\infty}S_n(k)\cos(k x-\omega t)d k\label{cov}
\end{equation}
is the normalized space-time covariance of the surface elevation~$\eta$ and $S_n(k)=S(k)/\sigma^2$. The Boccotti parameter $\psi^{*}$ is the first minimum of the time-covariance~$\psi(t)=\Psi(0,t)$ at $t=T^*$. Physically, $\zeta$ represents a dispersive broadbanded linear (small steepness) wave group, with no restriction on the spectral bandwidth. The largest crest occurs at the apex of the group as a result of a constructive interference, that is, the phases of its elementary Fourier components become perfectly coherent. Given a large crest amplitude $h\gg\sigma$, \cite{FEDELE2009} showed that the dimensionless residual $\Delta$ is Gaussian distributed with zero mean and standard deviation~$\sigma_{\Delta}=\sqrt{1-\psi^{*2}}\sigma/h$. As $h$ becomes larger, $\sigma_{\Delta}$ tends to zero and  $\Delta$ tends to a deterministic variable in the limit of infinite height~\citep{Boccotti2000,FEDELE2009}. 

In our simulations, we sample from the population of wave groups with  $\sigma_{\Delta}\approx0.1-0.2$ and sample values of $\Delta$ in $\pm\sigma_{\Delta}$. The large crest at $t=0$ is followed by a trough at $t=T^*+u$, with $u$ negligible as of $O(h^{-1}$~\citep{FEDELE2009}. From Eq.~\eqref{etac}, the trough amplitude is $h_{tr}=h(\psi^{*}+\Delta)$, where $\psi^{*}<0$~(see Figure~\ref{FIG1}). 

The parameter $\Delta$ measures the asymmetry of the wave profile.  In particular, the wave group~\eqref{etac} attains the given crest height $h$ at the focal point $(x=0,t=0)$, irrespective of $\Delta$, by way of a constructive interference of elementary waves. When $\Delta=0$, the phase coherence is perfect as the evolution of the wave group is perfectly symmetric before and after focus (see thin line in Figure~\ref{FIG1}).  Clearly, the probability of such perfectly symmetric occurrences ($\Delta=0$) is null as it occurs in the limit of infinite crest heights~\citep{Boccotti2000,FEDELE2009}. For a finite and large amplitude $h$, $\Delta\neq0$  breaks the symmetry of the stochastic wave group kinematics~(see bold line in Figure~\ref{FIG1}). This introduces hysteresis as explained below. 



The crest speed $c$ of the tallest crest of the stochastic wave group in Eq.~\eqref{etac}, where $\partial_{x}\zeta=0$, is given by~\citep{Longuet-Higgins1957}
\begin{equation}
c=-\frac{\partial_{xt}\zeta}{\partial_{xx}\zeta},\label{c}
\end{equation}
which is also valid for nonlinear surfaces. Stationarity is attained at the focusing point and the associated speed value  $c_s$ of $c$ is given by the weighted average of the linear phase speed $C(k)=\omega(k)/k$ of Fourier waves as~\citep{Fedele2014,Fedele2014a}
\begin{equation}
c_{s}=\frac{\int S(k)k^{2}C(k)dk}{\int S(k)k^{2}dk}+O(h^{-1}).\label{cmin}
\end{equation}
For spectra with narrow bandwidth~($\nu\ll1$),
\begin{equation}
c_{s}=c_{0}+\left[2k_{0}C'(k_{0})+\frac{1}{2}k_{0}^{2}C''(k_{0})\right]\nu^{2}+o(\nu^{2}),\label{css}
\end{equation}
where $c_{0}=C(k_0)$ is the linear phase speed at the spectral peak~$k=k_0$,  $C'(k_{0})$ and $C''(k_{0})$ are the derivatives of $C(k)$ with respect to the wavenumber $k$ evaluated at the same spectral peak. 

Moreover, from Eq.~\eqref{cmin}, following the crest path where $\partial_x\eta=0$~\citep{Fedele2014}, the time-varying crest speed $c(t)$ can be Taylor-expanded around the focusing time $t=0$ near the crest maximum
\begin{equation}
c=c_s+\frac{1}{2}\ddot{c}(0)t^{2}+O(t^4),
\end{equation}
and the time curvature at $t=0$ follows by taking the second derivative of~Eq.~\eqref{c} with respect to time
\begin{equation}
    \ddot{c}(0)=-\frac{c_{s}^{3}}{\partial_{xx}\zeta}\mathcal{L}(\partial_{x}\zeta),
\end{equation}
where the differential operator
\begin{equation}
    \mathcal{L}=\left(\partial_{x}+\frac{1}{c_{s}}\partial_{t}\right)^{3}.
\end{equation}
In spectral form
\begin{equation}
    \ddot{c}(0)=c_{s}^{3}\frac{\int k^{4}\left(1-\frac{C(k)}{c_{s}}\right)^{3}S(k)dk}{\int k^{2}S(k)dk}
\end{equation}
The sign of the time curvature $\ddot{c}(0)$ determines the tendency of the crest to slow down or speed up as the focal point is reached. In particular,
\begin{equation}
    \ddot{c}(0)=6k_{0}^{2}C'(k_{0})^{3}\nu^{2}+o(\nu^{2}).
\end{equation}

For linear dispersive waves with~$\omega(k)\sim k^n$, $C(k)\sim k^{n-1}$ and~$C'(k_0)=(n-1)k_{0}^{n-2}$. Thus, crests slow down in the deep-water wave regime since~$n=1/2<1$. On the contrary, crests speed up in the capillary wave regime, for which~$n=3/2>1$. Moreover, the slowdown/speedup is of~$O(\nu^{2})$. Thus, it arises with the modulation of the wave group and it increases as the spectrum becomes more broadbanded, as $\nu$ increases.


We note in passing that the above formalism and analysis for unidirectional (2D) wave groups are easily extendable to multidirectional (3D) wave packets drawing on~\cite{FEDELE2009}. In particular, consider a polar frame $(k,\theta)$ and the dominant direction of waves along $x$, or $\theta=0$. Then, the crest speed $c_s$ in Eq.~\eqref{cmin} extends to 3D groups as
\begin{equation}
c_{s, 3D}=\frac{\iint S(k,\theta)\,\omega(k)k\,\cos\theta\,dkd\theta}{\iint S(k,\theta)k^{2}\cos^{2}\theta\,dkd\theta}+O(h^{-1}),
\end{equation}
where $S(k,\theta)$ is the directional wave spectrum. Drawing on~\cite{fedele2015kurtosis}, consider narrowband waves with spectral bandwidth~$\nu$ and directional spreading~$\sigma_{\theta}$ around the dominant wave direction $\theta=0$. Then,
\begin{equation}
c_{s,3D}=c_{s,2D}\left(1+\sigma_{\theta}^{2}/2\right)+O(\sigma_{\theta}^{4}).
\end{equation}
Clearly, the crest slowdown of 3D wave groups manifests similar trends and properties as those of 2D wave groups, with a mild influence by the directional structure of the group. 

In summary, our analysis reveals that the crest slowdown effect in dispersive gravity-dominated waves can be very significant even during linear wave-focusing events, provided that the spectral bandwidth is sufficiently large. On the other hand, crest slowdown reduces as crests propagate over shallower waters, because waves tend to be less dispersive. In sharp contrast, crests of dispersive capillary-dominated waves tend to speed up. Finally,  the crest slowdown of 3D wave groups has similar behavior to that of 2D groups, as also clearly illustrated below by our results drawn from open ocean space-time data~(see Figure~\ref{FIG4F}). 

\begin{figure}
\centering\includegraphics[scale=0.42]{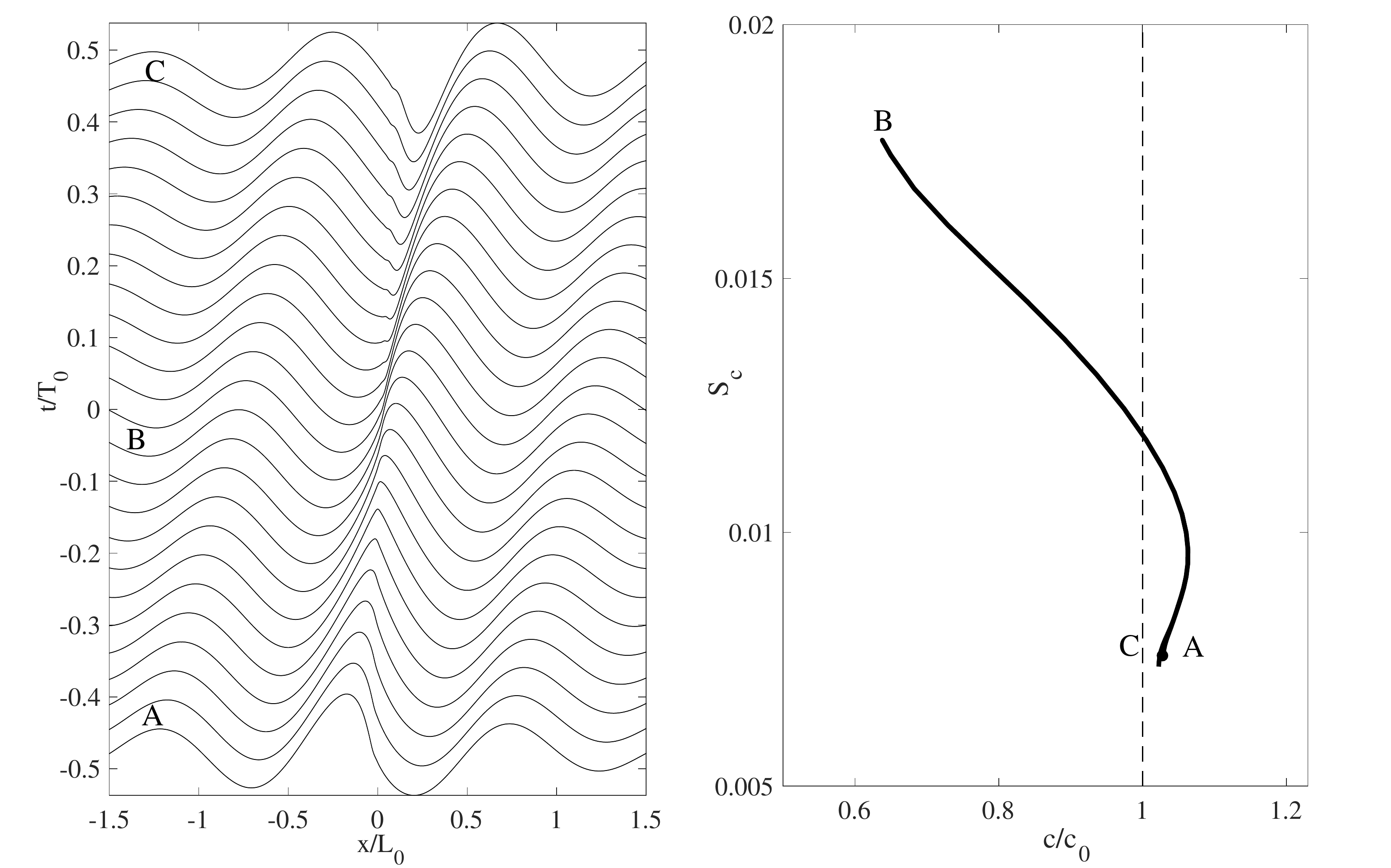} 
\caption{Perfect focusing of a deterministic linear broadband gravity wave group in deep water (JONSWAP spectrum, spectral bandwidth~$\nu\approx0.5$). Left panel: time stack of the surface profile $\zeta$ as waves propagate from left to right. Before focus (A), the crest leans forward as it slows down reaching the largest height at focus (B) and then leans backward while accelerating after focus (C). Note that the smaller adjacent crests undergo a similar evolution. Right panel: hysteresis curve of the normalized crest speed $c/c_0$ (bold line) of the tallest wave as a function of the local crest steepness $S_c$. Stokes theory predicts the crest speed $c/ c_0 =1$, as shown by the dashed vertical line. Here, as defined in Section~\ref{2_2}, $c_0$ is the reference linear deep-water phase speed of the dominant frequency~$\omega_0$. The leaning cycle has a duration close to the dominant period $T_0=2\pi/\omega_0$. }
\label{FIGF}
\end{figure}

\begin{figure}
\centering\includegraphics[scale=0.42]{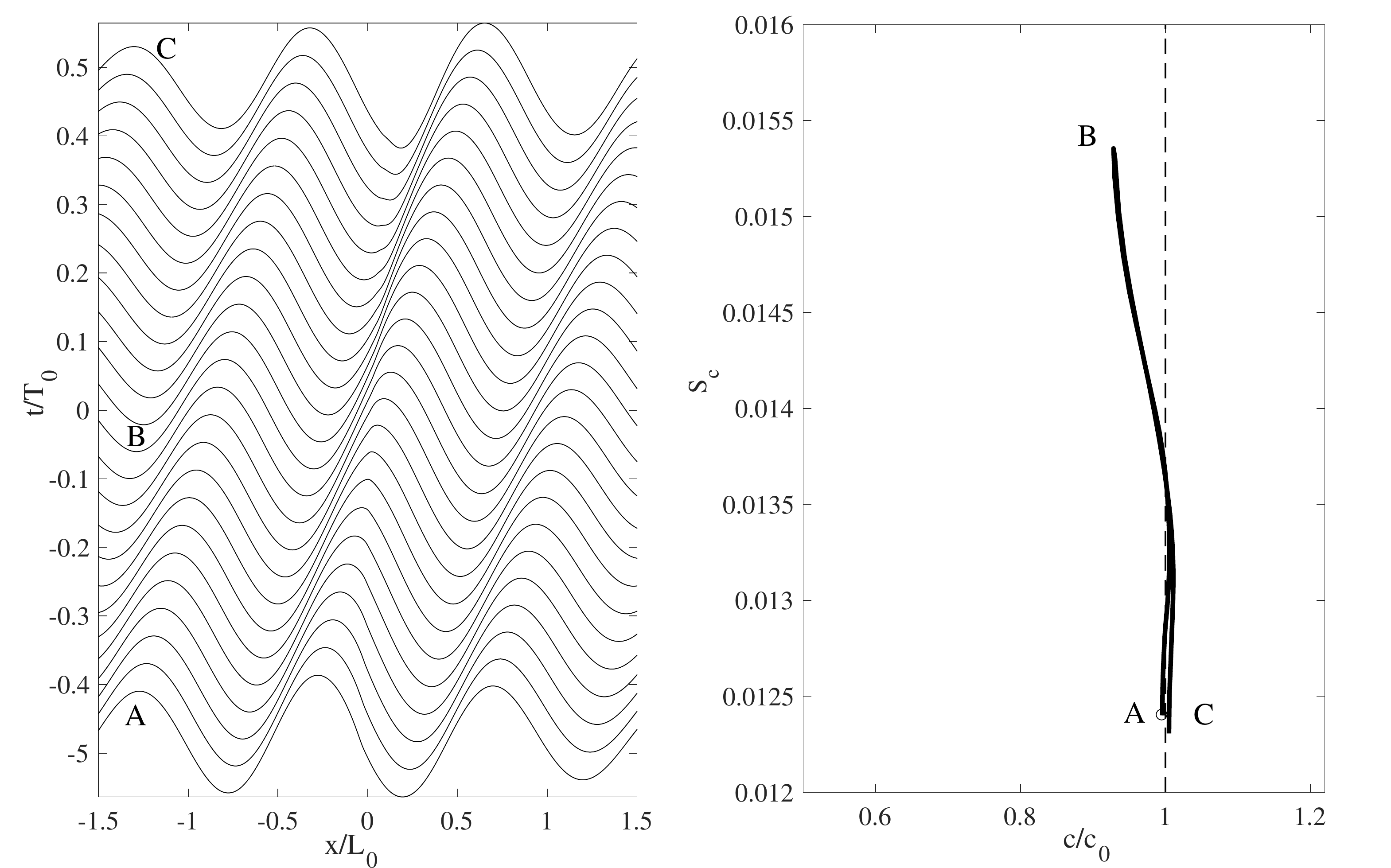} 
\caption{Focusing of the deterministic linear narrowband wave group in deep waters (JONSWAP spectrum, spectral bandwidth $\nu\approx0.1$). Other symbols are as in caption of Fig.~\ref{FIGF}.}
\label{FIG2F}
\end{figure}

\subsection{Crest slowdown of deep water gravity waves}\label{2_2}

Consider a unidirectional sea state of deep-water gravity waves with linear dispersion $\omega^2=g k$ and $g$ the gravity constant. The sea spectrum has a broad range of spectral frequencies (bandwidth $\nu\approx0.5$) and the conventional JONSWAP form. In particular, a peak enhancement parameter~$\gamma=3.3$ and spectral decay~$E(\omega)\sim \omega^{-5}$ or $S(k)\sim k^{-3}$, characteristic of realistic ocean waves~\citep{hasselmann1973,Komen1994}. We note in passing that our results and conclusions are the same for a Kolmogorov-Zakharov spectral decay~$E(\omega)\sim \omega^{-4}$~\citep{Zakharov1968,Zakharov1999}. We also assume a perfect (symmetric) focusing of the stochastic wave group in Eq.~\eqref{etac}, i.e. $\Delta=0$. Eq.~\eqref{etac} is solved numerically by standard Fourier methods ($N\sim10^4$ modes) on a computational domain long enough ($O(50)$ dominant wavelenghs $L_0$ ) to ensure zero-boundary conditions. The wave packets are localized in space and never reach the boundaries during the one-time focusing event.

The left panel of Figure~\ref{FIGF} depicts the time stack of the simulated surface profile $\eta$ as waves propagate from left to right. Significant leaning and slowdown are observed around a point of maximum crest height. In particular, the tallest wave crest of the group leans forward (A) prior to approaching the maximum of the group at the focusing site (B). At this stage, the crest has a steeper front face than its rear face, suggesting that it should accelerate as it approaches the focusing site and then eventually break. That would be the case if the wave were not able to continue to grow in amplitude. On the contrary, it is seen that the forward-leaning wave relaxes at (A) to a symmetrical shape as it accumulates potential energy reaching its tallest height at (B). While doing so, the wave crest decelerates significantly by~$O(30\%)$, as it attains its maximum height at (B). Since the process is linear, with perfect constructive interference, the reverse of the wave growth phase commences and it is exactly symmetrical with the earlier growing phase. After focus, the wave crest leans backwards at (C) and the rear face of the wave crest is now steeper than its front face. This suggests that the crest should now decelerate as it travels away from the focusing site. On the contrary, the backward-leaning wave crest face (C) accelerates while it reduces in amplitude, releasing the accumulated potential energy to the wave group. 

Note that the smaller adjacent crests undergo similar evolution. We observe that the leaning cycle has a duration close to the dominant wave period $T_0=2\pi/\omega_0$ irrespective of the spectral bandwidth. This is reasonable since all the waves in a group, irrespective of its length, undergo the same cyclical crest slowdown as also observed in our nonlinear numerical simulations shown below. 

Crest slowdown is more evident in the right panel of the same figure, which depicts the normalised crest speed $c/c_0$ of the tallest wave as a function of its local crest steepness $S_c=k_c a_c$. Here, $a_c$ is the amplitude of the crest and $k_c=2\pi/L_c$ is the associated local wavenumber, with $L_c$ being the distance between the two zero-crossings before and after the crest. The reference $c_0=\omega_0/k_0$ is the phase speed at the dominant frequency $\omega_0$. Note that Stokes theory predicts the crest speed $c/ c_0 =1$, as shown by the dashed vertical line.  Since the focusing is perfect, there is no hysteresis in the plot of $c/c_0$ versus $S_c$, since the growing and decaying phases of the group are symmetrical.  We point out that both the vertically-integrated potential and kinetic energy densities (hereafter as $PE$ and $KE$) following the crest of the tallest wave increase as the crest slows down, with the $PE$ growing more than the $KE$ since the $PE$ accumulates around the growing crest~\citep{Barthelemy2018}. Clearly, this is due to the linear nature of the wave group. Nonlinearities suppress the growth of potential energy while favouring that of the kinetic energy as a precursor to breaking. 

On the contrary, the same JONSWAP wave group with a narrowband spectrum~($\nu\approx0.1$) manifests a very mild crest slowdown and the changing shape of the crest form due to leaning is barely noticeable. This is clearly seen in Figure~\ref{FIG2F}. The life cycle of the crest is still symmetric and no hysteresis. 

\begin{figure}[h]
\centering\includegraphics[scale=0.40]{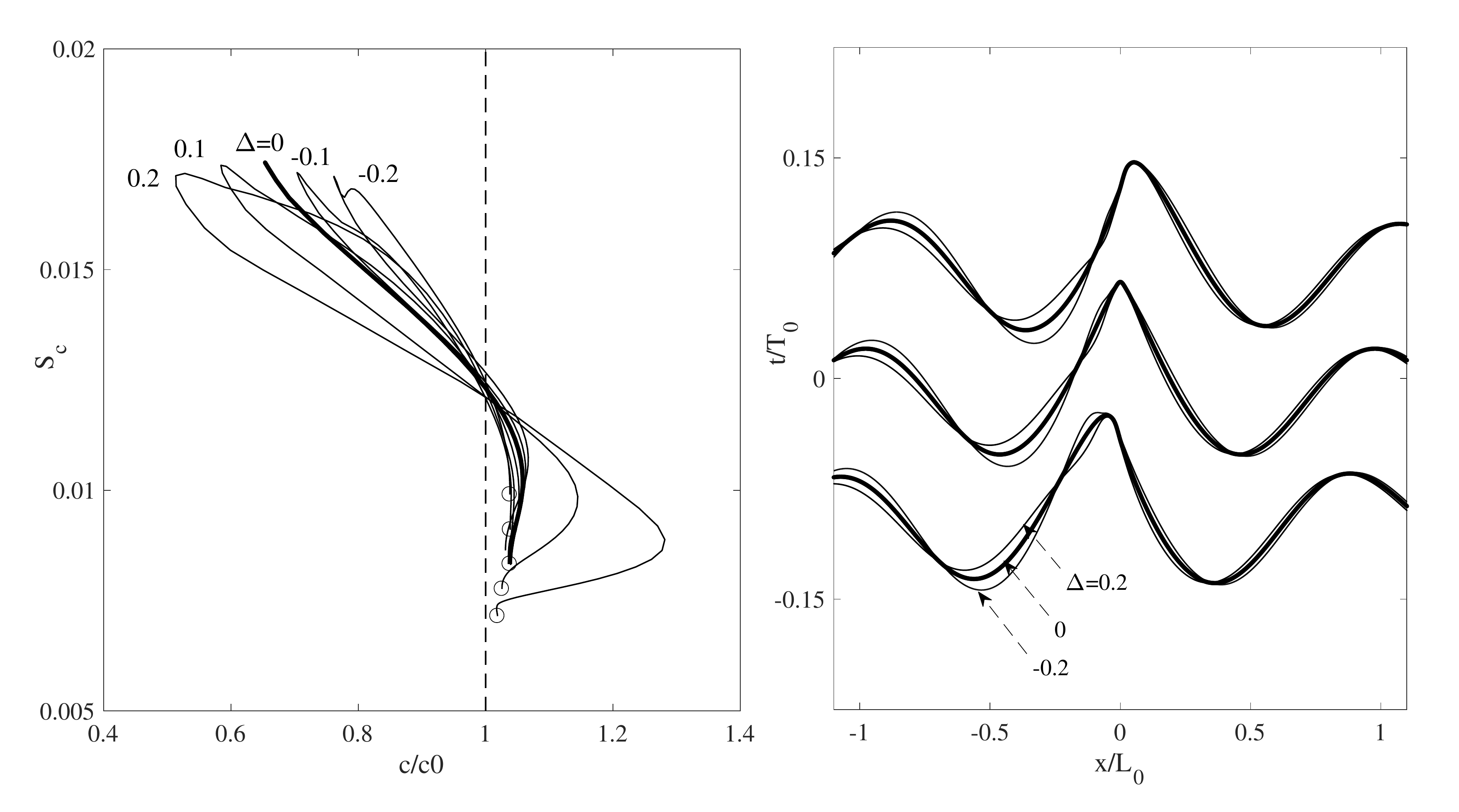} 
\caption{Linear crest slowdown and leaning cycle of a focusing gravity wave group in deep water (JONSWAP spectrum, spectral bandwidth $\nu=0.4$). Left panel: hysteresis curve of the normalized crest speed $c/c_0$ of the tallest wave as a function of its local crest steepness $S_c$ for different values of the parameter $\Delta$. Here, $c_0$ is the reference linear deep-water phase speed of the dominant frequency~$\omega_0$. Linear Stokes theory predicts the crest speed $c/ c_0 =1$, as shown by the dashed vertical line. Right panel: time stack of the surface profile $\zeta$ as waves propagate from left to right for three different values of $\Delta$, i.e. dash-dotted~($-0.2$), bold~($0$) and thin~($0.2$) lines. The leaning cycle has a duration close to the dominant period $T_0$. The crest leans forward as it slows down reaching the largest height at focus and then leans backward while accelerating.}
\label{FIGCF}
\end{figure}

Clearly, the probability of such perfectly symmetric occurrences ($\Delta=0$) is null as it occurs in the limit of infinite crest heights~\citep{Boccotti2000,FEDELE2009}. For finite and large amplitude $h$, the effects of the parameter $\Delta$ on kinematics of the stochastic wave group $\zeta$ and associated crest slowdown and leaning cycle are shown in Figure~\ref{FIGCF}. Here, we see that the crest height is unchanged  and we only observe variations in the leaning cycle that affect the crest slowdown. In particular, the slowdown increases as the crest tends to lean more backward (thin line, $\Delta>0$) in comparison with the symmetrical case (bold line, $\Delta=0$). On the contrary, the slowdown reduces as the crest leans less backward (dash-dotted line, $\Delta<0$). Moreover, the symmetry around focus is broken as clearly seen in Figure~\ref{FIGCF} from the diagrams of the crest speed $c/c_0$ versus the local crest steepness $S_c$. The curves enclose a finite area, which is a measure of the deviation from the ideal symmetry that characterizes perfect focusing. 

\subsection{Crest slowdown of gravity waves in intermediate depth water}

\begin{figure}[h]
\centering\includegraphics[scale=0.37]{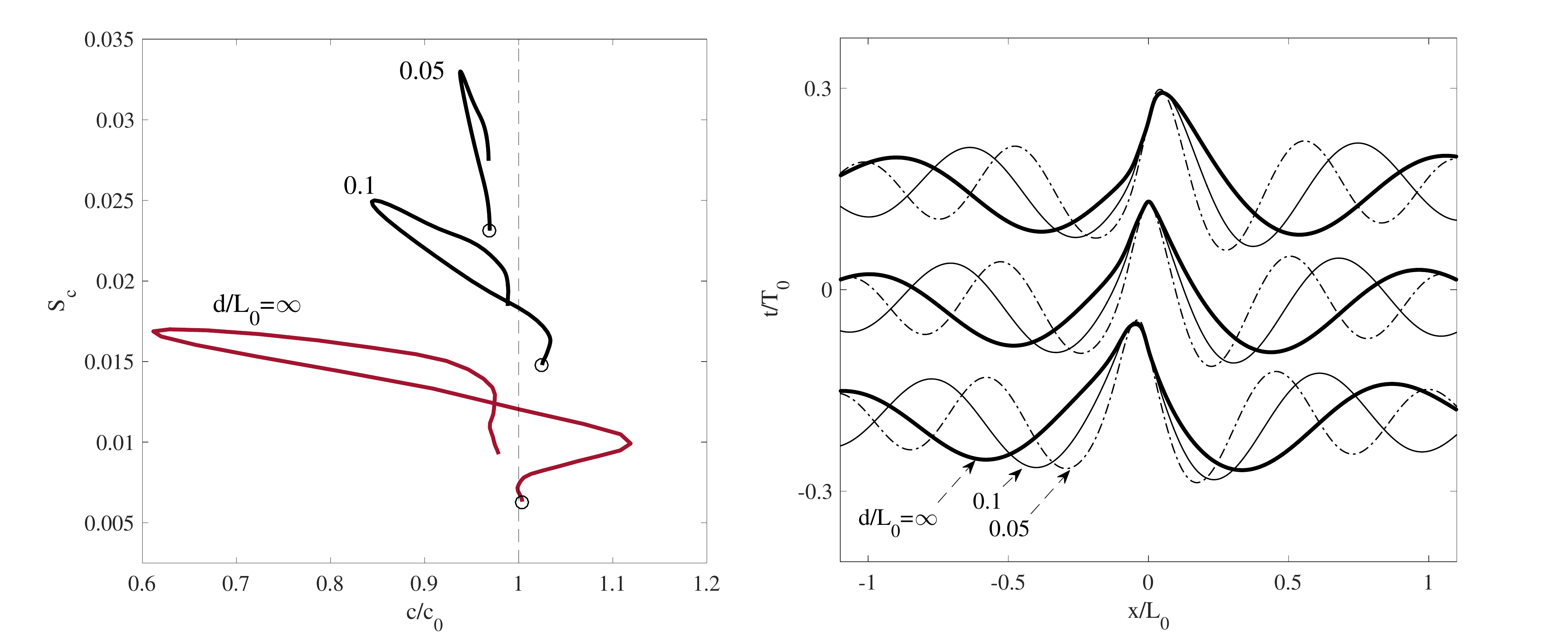} 
\caption{Water depth effects on the linear crest slowdown and leaning cycle of a stochastic wave group (JONSWAP spectrum, spectral bandwidth $\nu=0.3$, $\Delta=0.2$). Left panel: hysteresis curve of the normalized crest speed $c/c_0$ of the tallest wave as a function of the local steepness $S_c$ for different values of the water depth $d/L_0=0.05,0.1,\infty$. Here, $c_0$ is the linear phase speed at the finite depth $d/L_0$ and at spectral peak $\omega_0=2\pi/T_0$, with $L_0$ the associated wavelength. Right panel: time stack of the surface profile $\eta_c$ as waves propagate from left to right for the three values of $d/L_0$.}
\label{FIGDF}
\end{figure}

Consider a synthesized sea state of gravity waves propagating in waters of depth $d$ and linear dispersion relation $\omega^2=g k\tanh(k d)$. For simplicity we assume a JONSWAP spectrum for the evolution of the stochastic wave group in Eq.~\eqref{etac}, and parameters $\nu=0.3$, $\Delta=0.2$. Our results and conclusions are qualitatively the same for a Gaussian-shape spectrum. The left panel of Figure~\ref{FIGDF} depicts the hysteresis curves of the normalized crest speed $c/c_0$ of the tallest wave as a function of its local crest steepness $S_c$ for three values of water depth $d/L_0=0.05,0.1,\infty$. Here, we recall that $c_0$ is the linear phase speed for the depth $d/L_0$ and the spectral peak frequency $\omega_0=2\pi/T_0$, with $L_0$ the associated wavelength.

When the tallest wave of the group propagates over shallower water depths its local crest steepness $S_c$ increases, but the hysteresis cycles shift to the right and the relative slowdown within the cycle is also reduced. This reduction is associated with the fact that waves lean less when they propagate over shallower water depths, as clearly seen in the right panel of the same figure. In sharp contrast, in the following we will show that crests speed up in the capillary wave regime.


\subsection{Crest speedup in the capillary regime}

\begin{figure}[h]
\centering\includegraphics[scale=0.45]{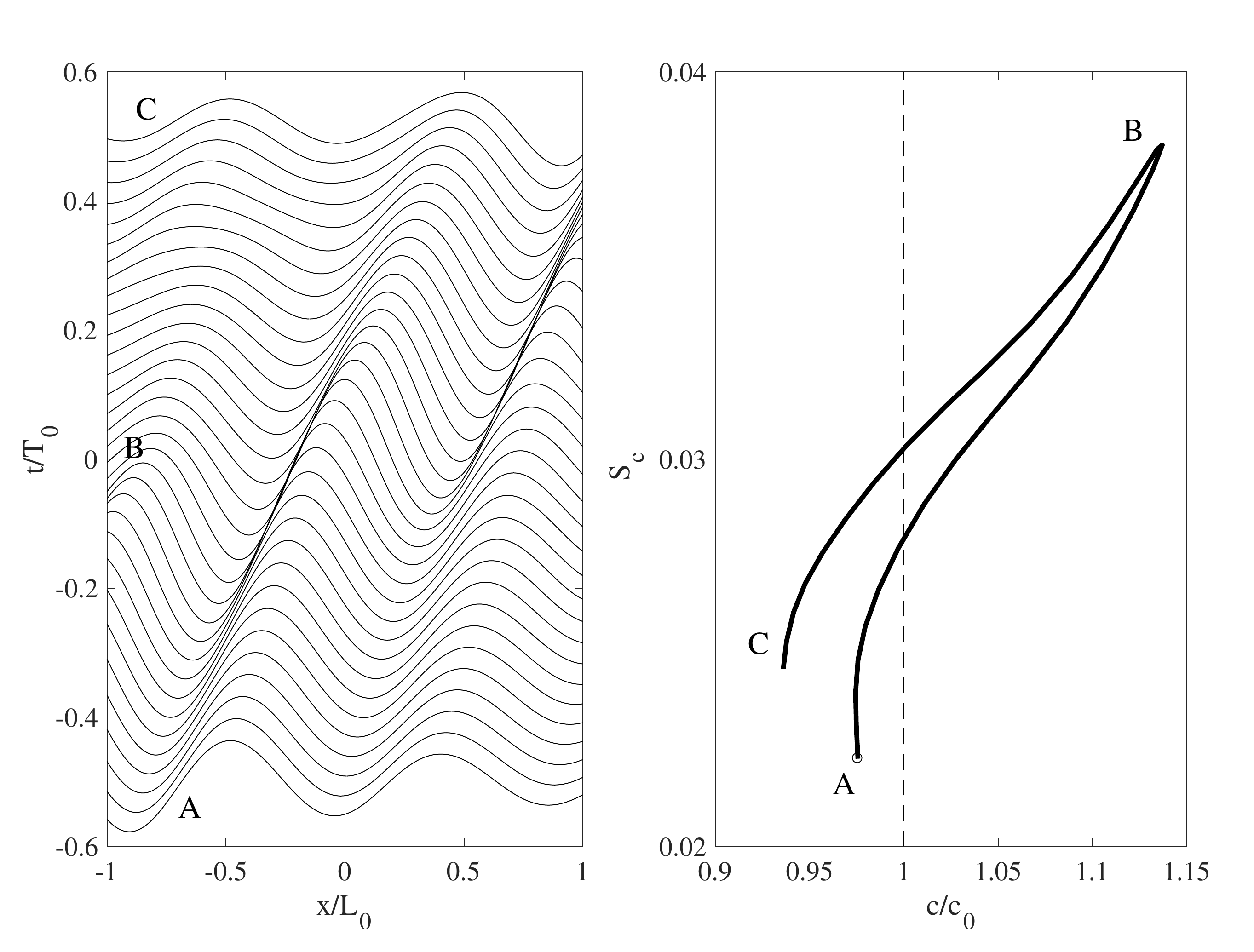} 
\caption{Capillary waves: focusing of a stochastic wave group. Here, $c_0$ is the reference linear phase speed of the dominant frequency~$\omega_0$. Other symbols are as in the caption of Fig.~\ref{FIGF}.}
\label{FIGDFca}
\end{figure}

Consider now the evolution of the stochastic wave group~$\zeta$ in Eq.~\eqref{etac} with capillary dispersion $\omega(k)\sim k^{3/2}$, and spectrum $E(\omega)\sim \omega^{-2.5}$, or $S(k)\sim k^{-3.25}$. We also set $\Delta=0.2$.  The left panel of Figure~\ref{FIGDFca} depicts the time stack of the surface profile $\zeta$ as waves propagate from left to right. In particular, the tallest wave crest of the group leans backward (A) prior to approaching the maximum of the group at the focusing site (B). At this stage, the crest has a gentler sloping front face than its rear face. The rearward-leaning wave at (A) relaxes to a symmetrical shape as it accumulates potential energy reaching its tallest height at (B). While doing so, the wave crest accelerates significantly as it grows in amplitude, attaining its maximum height at (B). Then, the reverse of the wave growth phase commences. In particular, after focus the wave crest then leans forward (C) and the forward face of the wave is now steeper than its rear face. The forward-leaning wave crest (C) decelerates while it reduces in amplitude, releasing the accumulated potential energy to the wave group. The leaning cycle has a duration close to the dominant wave period $T_0$. The crest speedup can be clearly observed in the right panel of the same figure, which depicts the hysteresis diagram of $c/c_0$ versus $S_c$. The speedup is more pronounced as the energy of shorter capillary waves becomes comparable to that of longer waves. This is because shorter capillary waves travel faster than longer waves. For instance, the speedup is practically unnoticeable if the spectrum $E(\omega)$ decays as faster as $\omega^{-5}$, or $S(k)\sim k^{-7}$.

In the following, we will study the effect of imperfect focusing on the crest slowdown. The maximum crest height is clearly affected, but we will show that the crest slowdown is practically insensitive to the degree of phase coherence of a group. Thus, crest slowdown is purely an effect of the natural dispersion of waves. In particular, the crest slowdown and associated leaning reduces as the dispersion is reduced when linear waves travel over shallower water depth. In the following, we will show that this is a generic property of the crest slowdown. Moreover, even nonlinearities affect the slowdown of crests because waves become less dispersive as they steepen.

\begin{figure}
\centering\includegraphics[scale=0.40]{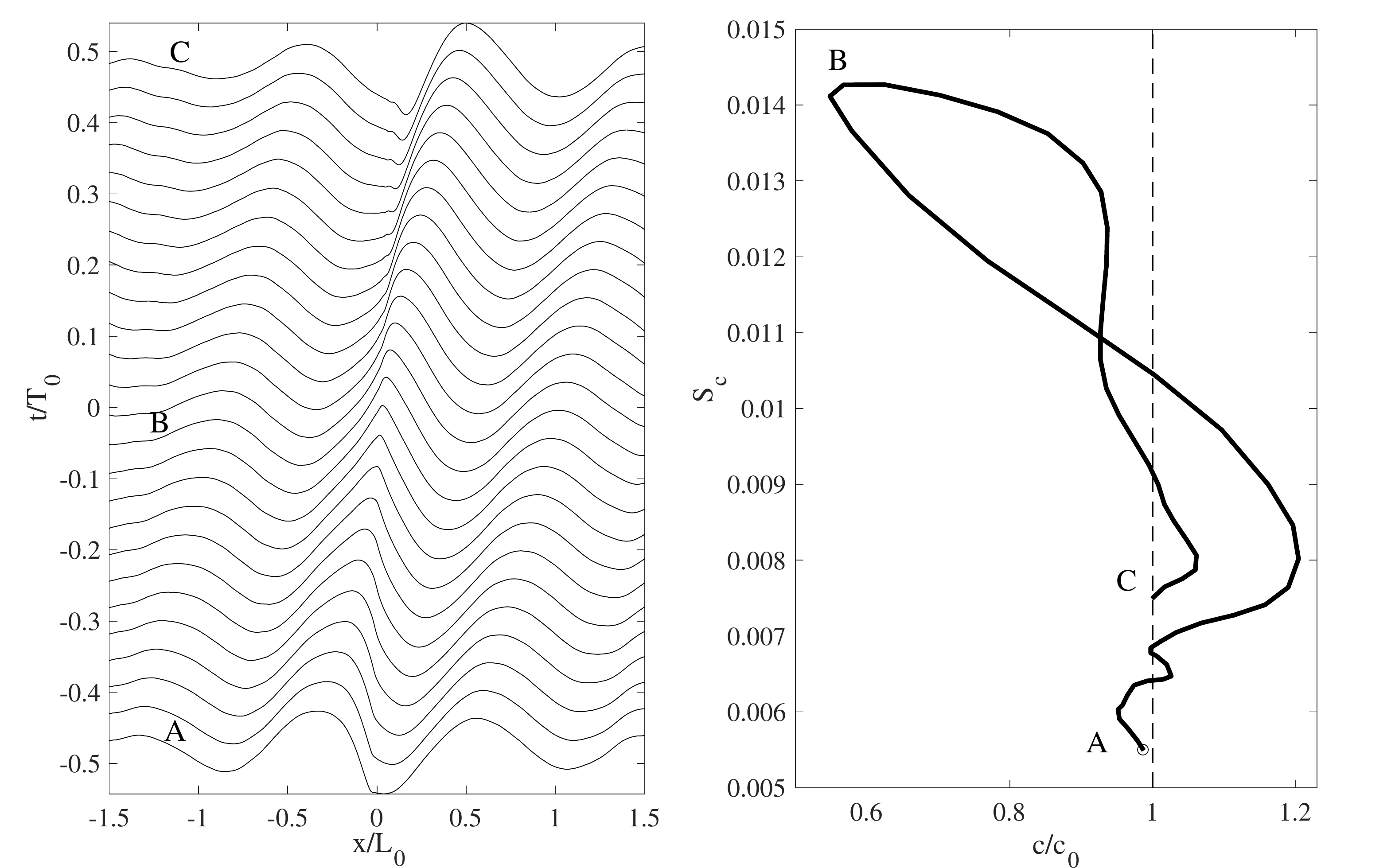} 
\caption{Linear crest slowdown of a focusing gravity wave group in deep water (JONSWAP spectrum, spectral bandwidth $\nu=0.4$, $\Delta=0.1$). Other symbols are as in caption of Fig.~\ref{FIGF}.}
\label{FIG3Fa}
\end{figure}

\subsection{Imperfect focusing}

Consider a linear wave group whose elementary Fourier components have phases not perfectly coherent. Drawing on~\cite{vanGroesen2014}, the associated surface displacements can be described by~\citep{Fedele2008,FEDELE2009} 
\begin{equation}
\zeta(x,t;\alpha)=h\int_0^{\infty}S_n(k)\cos(k x-\omega(k) t+\alpha \theta(k))d k.\label{etacr}
\end{equation}
When $\alpha=0$ (perfect coherence), the maximum crest height $h$ is attained at focus. 
The degree of phase coherence is measured by the defocussing factor $\alpha$~\citep{vanGroesen2014} and $\theta(k)$ is a random phase function of $k$. In particular, at any $k$, the phase $\theta$ is a uniform random variable distributed in $[-\pi,\pi]$ with mean $\mu(k)=0$ and standard deviation $\sigma(k)=\pi/\sqrt{3}\approx0.58\pi$. A set of $N$ phases $\theta_j=\theta(k_j)$ at different wavenumbers $k_j$ are stochastically independent and their joint probability density function (pdf) is
\begin{equation}
p(\theta_1,\theta_2,...\theta_N)=\frac{1}{(2\pi)^N}.\label{pdf}
\end{equation}
Note that this is one of the ways to model the imperfect phase coherence of a wave group. A first look at $\zeta$ in Eq.~\eqref{etacr} suggests that it is a random process. On the contrary, $\zeta$ is deterministic as a result of the strong law of large numbers. Therefore, it assumes the form~(see appendix A) 
\begin{equation}
\zeta(x,t;\alpha)=\widetilde{h}(\alpha)\int S_n(k) \cos(k x -\omega t) dk. 
\end{equation}
This is identical to a perfectly focusing packet $\zeta(x,t;\alpha=0)$ in Eq.~\eqref{etac}, except that now its crest amplitude is smaller than the maximal height $h$ as a consequence of phase decoherence, that is
\begin{equation}
\widetilde{h}(\alpha)=h\frac{\sin(\pi\alpha)}{\pi\alpha}.
\end{equation}
Nevertheless, the crest of the defocused wave group~($\alpha\neq0$) still experiences the same amount of linear slowdown and leaning as those of a perfectly focused wave group~($\alpha=0$), because the shape of the packet is unchanged, that is
\begin{equation}
 \frac{\zeta(x,t;\alpha)}{\widetilde{h}(\alpha)}=\frac{\zeta(x,t;0)}{\widetilde{h}(0)}. 
\end{equation}
As a result, from Eq.\eqref{c}
\begin{equation}
c(\alpha)=-\frac{\partial_{xt}\zeta(x,t;\alpha)}{\partial_{xx}\zeta(x,t;\alpha)},\label{cc}
\end{equation}
and, with probability 1, the minimum crest speed 
\begin{equation}
c_s(\alpha)=c_s(0),  
\end{equation}
implying that an imperfectly focused wave crest slows down as much as a perfectly focused wave crest, and the speed $c_s$ is given in Eq.~\eqref{cmin}.  In passing, we point out that 3D wave groups have similar behavior.





In summary, the crest slowdown of linear gravity wave groups is insensitive to the degree of phase coherence of the wave group. The associated leaning cycle is purely the effect of the natural dispersion of waves as discussed later on. Note also that nonlinear effects on the crest slowdown are attenuated since the wave amplitude is reduced.



\section{Geometry and nonlinearities} 
Geometric changes of each individual wave inside the packet are strongly phase-locked as seen, for example, in Figure~\ref{FIG3Fa}. Each wave leans forward following its creation at the rear of the packet, becomes symmetric as it reaches the crest maximum at focus and then leans backward as it disappears at the front of the packet. The leaning or asymmetry of the evolving shape of a focused crest is purely a linear phenomenon due to the dispersive nature of water waves, and it can be explained in terms of geometric phases~\citep{Fedele2014}.  

The forward leaning of the wave before focus~(sharper front and gentler back), is the kinematic manifestation of the longer wave components being behind the shorter ones. As focus is attained, longer waves catch up with the shorter ones. As a result, the wave accumulates potential energy as it slows down and the crest amplitude increases. At focus, the profile becomes symmetric and the crest attains its maximal height as the potential energy growth is zero. After focus, the wave crest accelerates while it reduces its amplitude, releasing potential energy to the wave group. The longer waves are now ahead of the shorter ones and the wave leans backward~(sharper back and gentler front). Thus, the leaning of crests is the kinematic manifestation of linear dispersive focusing. 


According to~\cite{Fedele2014}, the crest slowdown phenomenon, which manifests when unsteadiness plays a role in the wave group propagation, has its theoretical origin in the geometric phase framework in quantum mechanics (e.g.~\cite{Berry1988,Wilczek1989}). Only recently, it has been investigated in detail for unsteady deep water wave packets~\citep{Fedele2014}. The forward leaning and subsequent crest slowdown are primarily linear effects induced by the dispersive nature of waves. Briefly, a linear unsteady wave group propagates at a speed that includes a dynamical phase velocity dependent on the mean spectral frequency, and an additional geometric phase velocity induced by the changing shape of the crest due to leaning. \cite{Fedele2014} calculates that the leaning coefficient at crest maxima of unsteady oceanic groups attains a slowdown of~$c/c_0\approx 0.8$. Other fluid dynamical phenomena with the same underlying physics include self-propulsion at low Reynolds numbers~\citep{Wilczek1989} and the speed of fluctuating coherent structures in turbulent pipe flow that depends on both their dynamical inertia and also on their geometrical shape as it  varies in time~\citep{Fedele2015}.

Clearly, the unsteadiness of dispersive focusing is essential for crest slowdown to be observed. The slowdown of a crest is the manifestation of energy flowing through the wave packet: it accumulates in the slowing crest as potential energy, causing the growth in crest amplitude. Then, it is released into kinetic energy as the crest accelerates while reducing its height.  However, for waves in deep water, theoretical considerations suggest that third-order nonlinear dispersion tends to speed up the crest, quenching its growth~\citep{Fedele2014a}. In particular, consider the surface elevation~$\eta=\epsilon a_0 B(X,T) e^{i(k_0 x-\omega_0 t)}$ of a weakly nonlinear, broadband wave group in deep water. Here, $a_0$ is the amplitude of the carrier wave at frequency $\omega_0$ and wavenumber $k_0$, and $\epsilon$ is a small parameter to order nonlinearities in terms of spectral bandwidth. Here, the carrier wave is a sinusoidal waveform that is modulated or modified by both linear dispersion and nonlinearities.  The dimensionless surface envelope $B(X,T)$ varies on the slow space and time scales $X=\epsilon k_0 (x-C_g t)$ and $T=\epsilon^2 \omega_0 t$, where $C_g=c_0/2$ is the group velocity. The envelope satisfies, correct to $O(\epsilon^3)$, the compact Zakharov (cDZ) equation~\citep{dyazakh2017,dyazakh2017a} for unidirectional waves, which follows from the general Zakharov equation~\citep{Zakharov1968,Zakharov1999,krasitskii1994}. The essential form of the cDZ is that of a modified Nonlinear Schr\"odinger (NLS) equation~\citep{Fedele2012,Fedele2014a,FedeleJETP2014}
\begin{equation}
iB_{T}=\frac{1}{8}B_{XX}+\frac{i\epsilon}{16}B_{XXX}+\underset{NLS}{\underbrace{\left|B\right|^{2}B}}-\underset{Dysthe}{\underbrace{3i\epsilon\left|B\right|^{2}B_{X}}}-\underset{cDZ}{\underbrace{\epsilon^{2}\left|B\right|^{2}B_{XX}}}+\epsilon^3 N(B),\label{cDZ}
\end{equation}
where $O(\epsilon^3)$ terms have been collected in $N(B)$~\citep{Fedele2012}.
 Here, $B_T=\partial_T B$ and $B_X=\partial_X B$ denote the partial derivatives with respect to the slow time and space scales, respectively. The $O(\epsilon^0)$ nonlinear cubic term is that of the NLS equation. The $O(\epsilon)$ nonlinear term is that of the Hamiltonian Dysthe equation~\citep{Dysthe1979,Fedele2012,FedeleJETP2012}, and the new $O(\epsilon^2)$ term is characteristic of the cDZ equation. We note in passing that the original \cite{Dysthe1979} equation is expressed in terms of physical variables, which are non-canonical and the symplectic structure is lost. However, it can be recovered by way of a canonical transformation~\citep{Fedele2012,FedeleJETP2012}. All three nonlinear terms tend to reduce the dispersive nature of waves, especially the cDZ term, which is the most effective as the wave envelope $B$ grows in amplitude and broadens in spectral bandwidth during focusing. Indeed, the linear dispersive tendency of the wave packet, as modelled by~$B_{XX}$, mostly reduces by the factor~$(1-8\epsilon^2\left| B \right|^2)$ due to the cDZ term. This term allows for soliton solutions in the form of peakons with discontinuous slope~\citep{Fedele2012}. As a result, there is a nonlinear increase in the phase speed $C(k)$ of higher wavenumber Fourier components, which offsets the linear crest slowdown. To see this, Eq.~\eqref{cDZ} is rewritten in the compact differential form
\begin{equation}
\left\{ 1-i\epsilon/2\partial_{X}+i\epsilon^{2}\partial_{T}-\Omega\left(1-i\epsilon\partial_{X},\epsilon^{2}\left|B\right|^{2}\right)\right\} B=0
\end{equation}
where the associated nonlinear dispersion relation $\Omega(k,\left|B\right|^{2})$ is given by the Taylor expansion 
\begin{equation}
\Omega\left(k,\left|B\right|^{2}\right)=1+\frac{K}{2}-\frac{K^{2}}{8}+\frac{K^{3}}{16}+\left|B\right|^{2}\left(1+3 K+K^{2}\right),\label{OM}
\end{equation}
where $k=1+K$ and $k$ denotes dimensionless wavenumber (normalized by $k_0$), and the first four terms represent the third-order Taylor expansion of the deep-water frequency $\sqrt{1+K}$ with respect to $K$. The nonlinear frequency  $\Omega(k,\left|B\right|^{2})$ increases as the envelope amplitude $\left|B\right|$ increases. So does the minimum crest speed in Eq.~\eqref{cmin}, where the nonlinear phase speed $C(k)=c_0\Omega\left(k,\left|B\right|^{2}\right)/k$ in the integrand increases, leading to larger crest speeds.  Thus, the crest slowdown is reduced and the wave growth is suppressed~\citep{Fedele2014a,fedeleJFM2016}.

The Dysthe term does not suppress the NLS (subharmonic) modulational instability (MI), which is instead affected by the cDZ term. Indeed, MI is suppressed as the wave steepness $\mu=2\left|B\right|$ of the packet increases above $0.27$ and breathers are annihilated~\citep{Fedele2014a,FedeleJETP2014}. Further, the cDZ term induces superharmonic instability as $\mu$ exceeds the critical value $\mu_c=0.577$ and the envelope dynamics change from elliptic (NLS) to hyperbolic type. In particular,~\cite{Fedele2014a} showed that for $\mu>\mu_c$ a perturbation $\epsilon F(\xi,\tau)$ to the squared envelope $E=\left| B(X,T) \right|^2$ evolves according to the hyperbolic modified KdV equation
\begin{equation}
F_{\tau}+\beta F F_{\xi}+\epsilon\left(z_{1} F_{\xi\xi\xi}+z_{2}F^{2} F_{\xi}\right)+\epsilon^{2}z{}_{3}\left(2 F_{\xi}F_{\xi\xi}+ F F_{\xi\xi\xi}\right)=0,\label{KdV}
\end{equation}
where $\xi=\epsilon (X-cT)$, $\tau=\epsilon^{2} T$ are independent multiple scales, $c$ is the wave celerity, and the coefficients $(\beta,z_j)$ are given in~\cite{Fedele2014a}. Eq.~\eqref{KdV} typically arises in shallow water theory and it describes the tendency of the wave packet to steepen and eventually break in finite time. This suggests possible physical similarities between shallow- and deep-water waves. \cite{Fedele2014a} also suggests the transition from elliptic to hyperbolic behavior in the wave dynamics as the breaking onset, the point of no return beyond which waves inevitably break~\citep{Barthelemy2018}.

In summary, a key open question discussed in~\cite{Fedele2014,Fedele2014a,fedeleJFM2016} is concerned with the effect of wave nonlinearities on the crest slowdown. In the following, we present some theoretical arguments that provide some initial insights, suggesting that nonlinearities have a secondary effect on the magnitude of the leaning and consequently on the slowdown.

\subsection{Nonlinear dispersion effects on the crest slowdown}

We have seen that crest slowdown is a generic feature of linear gravity water wave systems, with relatively large slowdowns ($c/c_{0}<0.7$) obtained in our simulations. In comparison,  a more moderate slowdown $c/c_{0}\approx0.8$ was found in our present numerical simulations and in field observations~\citep{Banner2014}. The reduced slowdown can be explained by accounting for the nonlinear dispersive nature of waves. The slowdown of the largest crest in a wave group reduces because steeper waves are less dispersive at higher wave steepness. In particular, as the wave amplitude increases, shorter elementary waves increase their phase speed, catching up with their longer counterparts. As a result, the nonlinear increase in phase speed offsets the linear crest slowdown due to the finite spectral bandwidth, limiting the slowdown of the wave group. This can be seen from Eq.~\eqref{cmin} for the minimum crest speed, where the phase speed $C(k)$ in the integrand increases, leading to larger crest speeds.   

The hypothesis that the reduction of the crest slowdown is due to the nonlinear dispersive nature of waves~\citep{Fedele2014,Fedele2014a} is validated by way of recent oceanic space-time field observations using a Wave Acquisition Stereo System~(WASS, see~\cite{FedeleOMAE2011,Benetazzo2012,Gallego2011,Gallego2013,Fedele2013}) and hereafter discussed.

\begin{figure}[h]
\centering\includegraphics[scale=0.5]{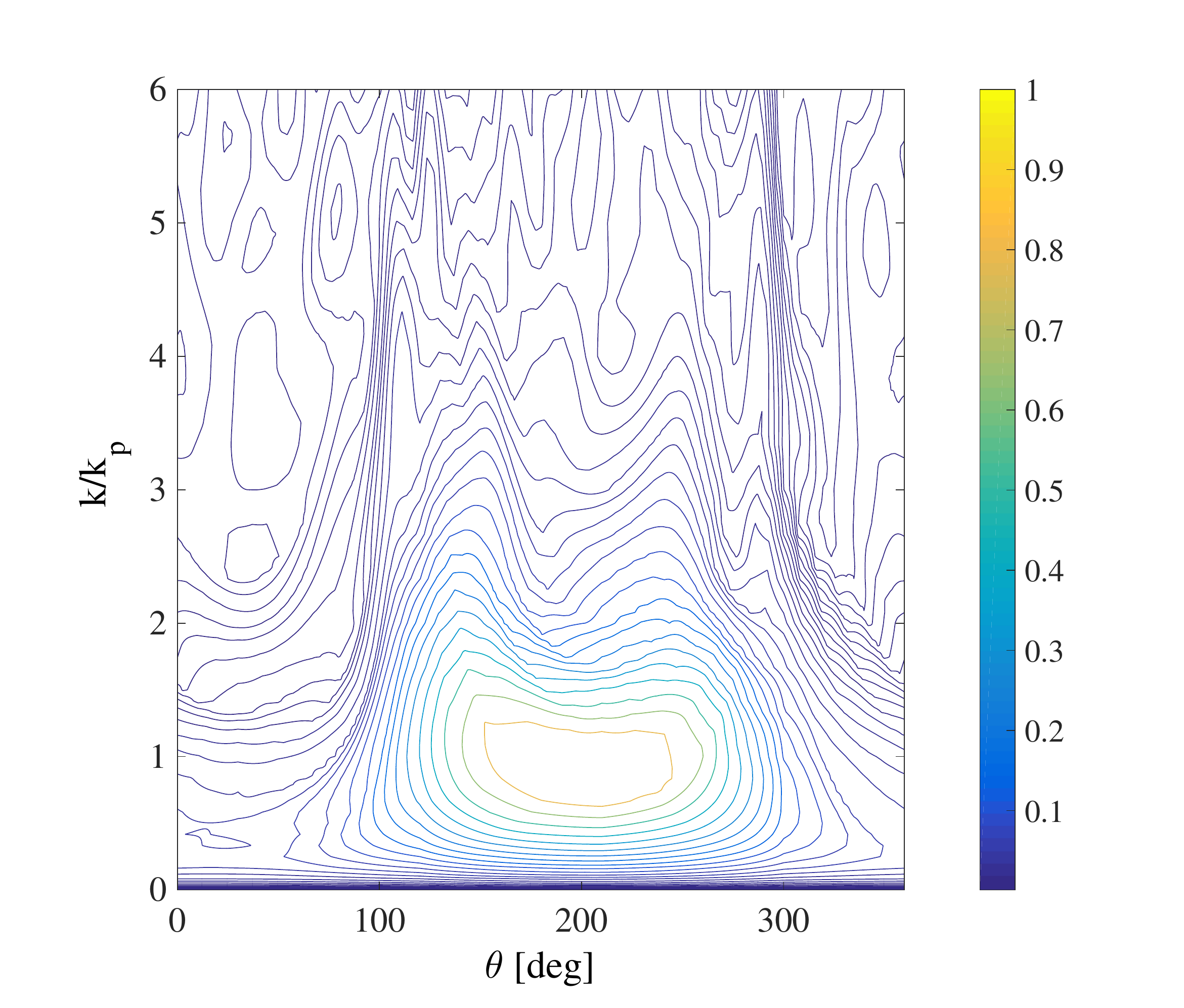} 
\caption{WASS observations at Acqua Alta: observed directional spectrum $S(k,\theta)/S_{max}$ and $k_p$ is the wavenumber at the spectral peak $S_{max}$.}
\label{FIGspec}
\end{figure}

\begin{figure}[h]
\centering\includegraphics[scale=0.45]{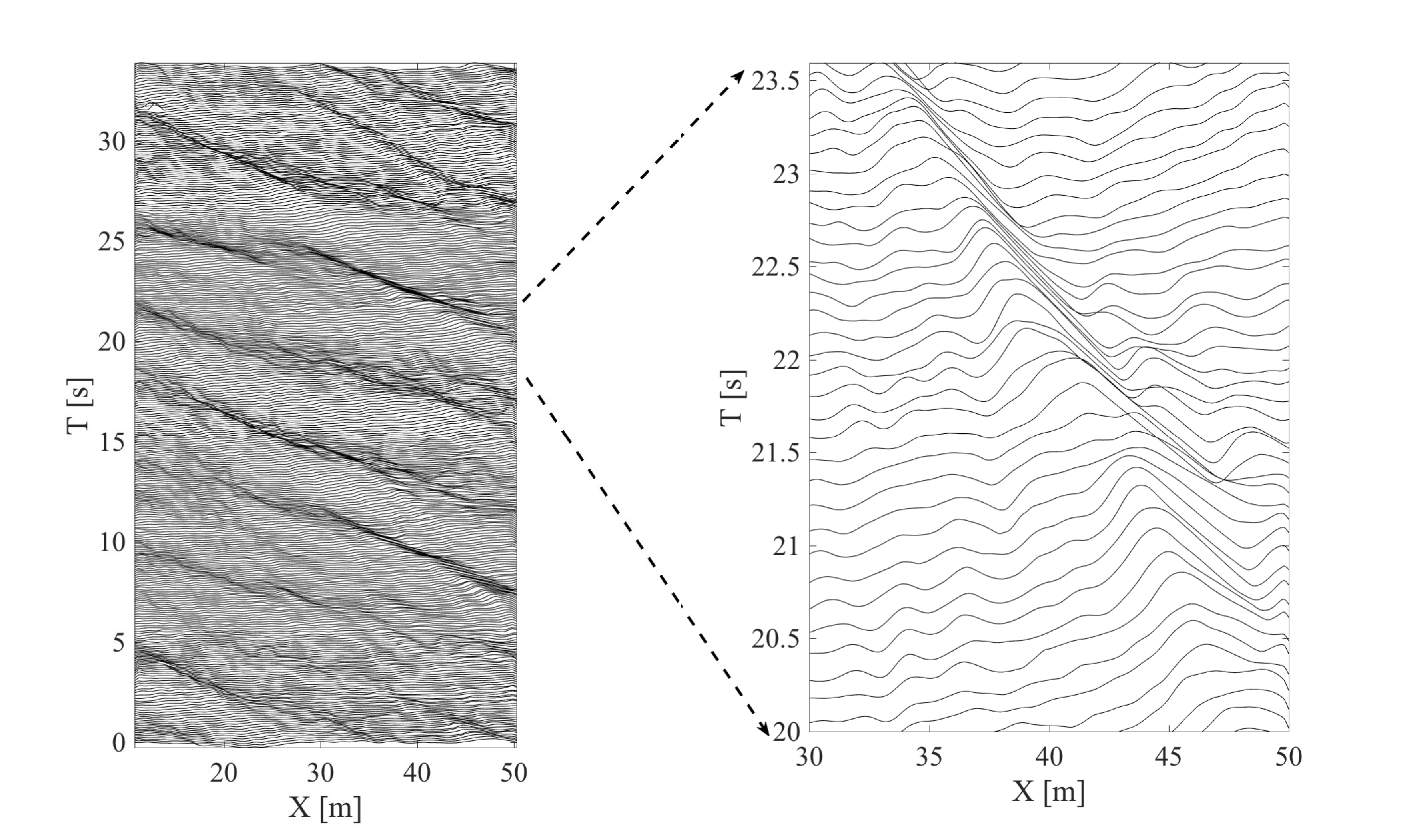} 
\caption{WASS observations at Acqua Alta: (left) Time stack of the stereo-observed wave field along the dominant direction (wave profiles stack at the sampling frequency $f_s=10$~Hz). Waves propagate from left to right; significant wave height $H_{s}=1.09$~m and zero-crossing period $T_{z}=3.51$~s; (right) details of the leaning cycle of a large crest that leans forward as it slows down before reaching the largest height and then leans backward as it accelerates. Note that the cycle time scale is about~$T_z$.}
\label{FIG3F}
\end{figure}

\begin{figure}[h]
\centering\includegraphics[scale=0.55]{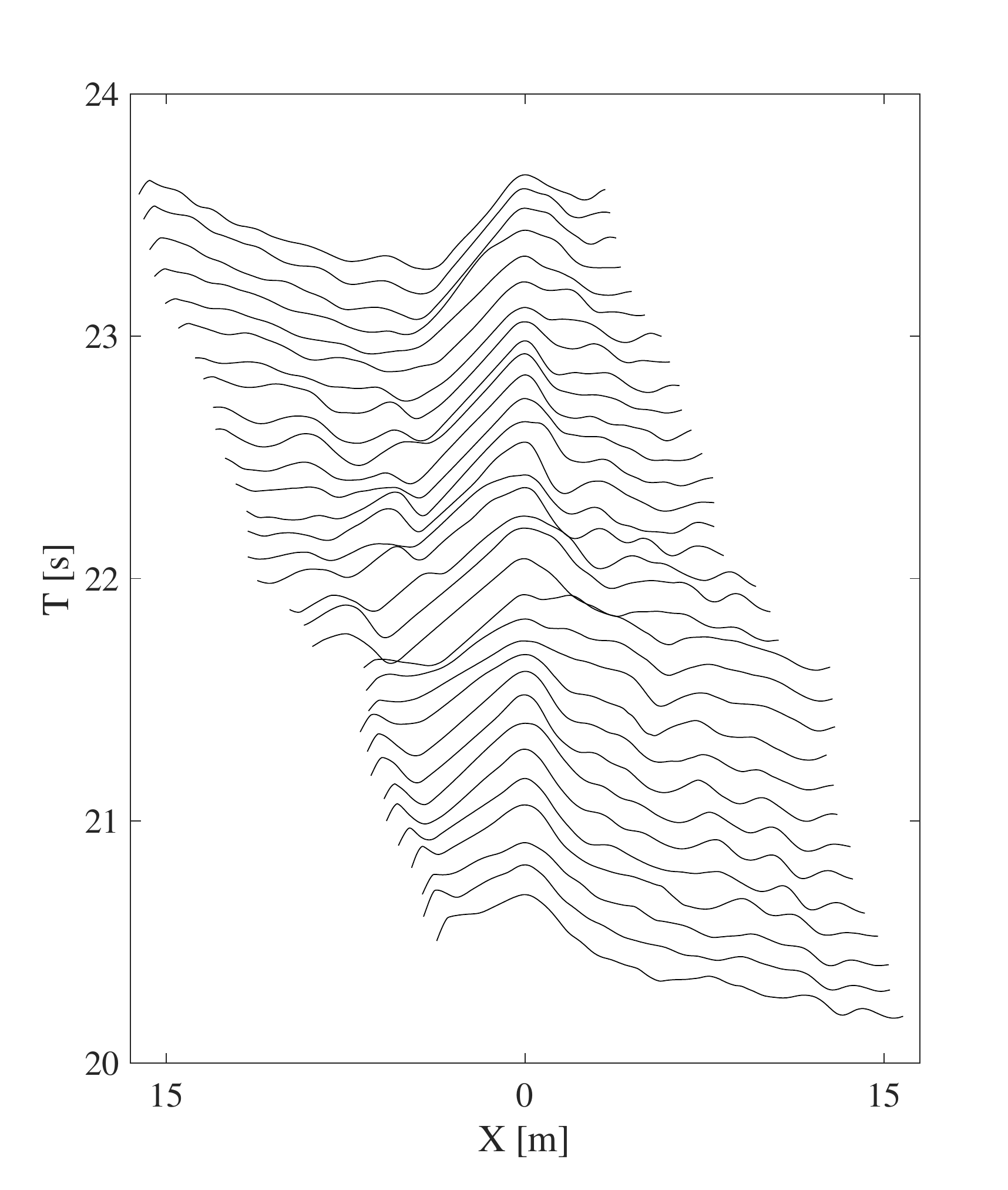} 
\caption{WASS observations at Acqua Alta:  leaning cycle of the large crest shown in Fig.~\ref{FIG3F} in a reference frame moving with the crest.}
\label{FIG4F}
\end{figure}

\begin{figure}[h]
\centering\includegraphics[scale=0.40]{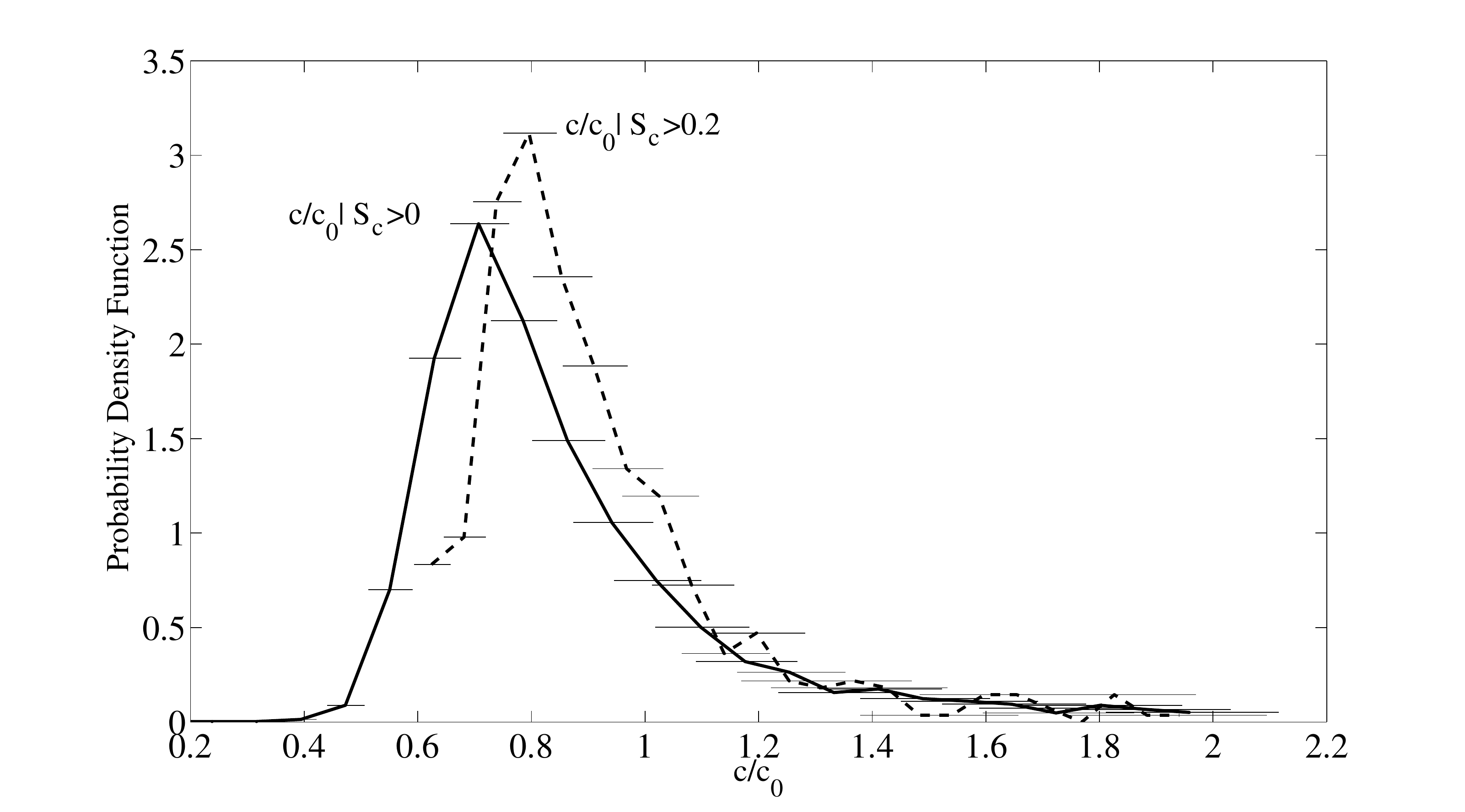} 
\caption{WASS observations at Acqua Alta: (solid line) marginal probability density function (PDF) of the normalized crest speed~$c/c_0$ and (dashed line) conditioned on crest steepness values~$S_c>0.2$.} 
\label{FIG1F}
\end{figure}

\subsection{Ocean field observations}

WASS was deployed at the oceanographic tower Acqua Alta in the Northern Adriatic Sea, 16 km off the coast of Venice in 16 m deep water. Stereo video measurements were acquired in three experiments during 2009-2010 to investigate space-time and spectral properties of ocean waves~\citep{FedeleOMAE2011,Gallego2011,Benetazzo2012,Gallego2013,Fedele2013}. To maximize the stereocamera overlap area, WASS was deployed $12.5$~m above the mean sea level at a $70^{o}$ depression angle. This provided a trapezoidal overlap area with sides of $30$~m and $100$~m, a width of $100$~m and an imaged area of approximately $1100$~$m^2$. 
We describe results from data acquired during Experiment 2, using 21,000 frames captured at $10$~Hz. The mean windspeed was $9.6$~m/s with a $110$~km fetch, and the unimodal wave spectrum had a significant wave height $H_s=1.09$~m and zero-crossing period $T_{z}=3.51$~s~\citep{FedeleOMAE2011,Fedele2013}. The observed directional spectrum $S(k,\theta)$ is shown in Figure~\ref{FIGspec}. Note the bimodal structure of the spectrum in the high wavenumber range as observed in oceanic sea states~\citep{BannerYoung1994,Young1995,Ewans1998,Hwang2000,Hwang2001,ToffoliJGR2010}. Nevertheless, the imaged dominant crests still undergo a slowdown cycle similar to the simulated crests described above.

Analyses performed include estimating the speed $c$ of focusing crests that travel through the imaged area. Most of the observed crests are very steep and close to breaking. The data were filtered above $1.5$~Hz to remove short riding waves. The speeds of crests observed within the imaged area were estimated by tracking the spatial crest along the dominant wave direction. Subpixeling reduced quantization errors in estimating the local crest position. As the imaged area was not large enough to capture a complete focusing event, the local reference wave speed $c_0$ was calculated using the linear dispersion relationship from the spectral peak frequency of a short time series centered at the crest ($D=120$~s as an optimal record length) and Doppler corrected for the in-line $0.20$~m/s mean current. A local wavelength $L_c$ for each crest was estimated as the distance between the two zero-crossings adjacent to the  crest. Then, the local crest steepness  was estimated as $S_c=k_c a_c$, where $k_c=2\pi/L_c$ is the local wavenumber and $a_c$ the crest height.

The left panel of Figure~\ref{FIG3F} shows the space-time evolution of the stereo-reconstructed wave field along the dominant direction. Waves propagate from left to right. In the right panel of the same figure, a detailed evolution of one of large crests shows that the crest leans forward as it decelerates before reaching the largest height and then it speeds up as it leans backward. This is clearly seen in a frame moving with the crest, as shown in Figure~\ref{FIG4F}. 
 
Moreover, the marginal probability density function of the normalized crest speed $c$ by itself estimated from all the measured crests peaks at close to $0.8c_0$ as is clearly seen in Figure~\ref{FIG1F}. However, the same speed distribution conditioned on values of wave steepness $S_c>0.2$ peaks at $0.9c_0$ indicating a reduced slowdown. This is consistent with the nonlinear dispersive nature of waves as discussed above. 
In the following section, we show that these ocean observations are also consistent with our numerical simulations of unsteady evolution of unidirectional (2D) and multidirectional (3D) dispersive wave packets in both deep and intermediate depth waters, for a range of wave steepness levels. A reduced slowdown is also observed for waves travelling in intermediate water depth, where waves tend to be less dispersive because of the influence of the ocean bottom.

\section{Numerical results}

To investigate crest speed slowdown of modulating steep gravity waves, our present study used results from simulations using the boundary element numerical wave tank code WSIM, a 3D extension of the 2D code developed by~\citet{Grilli1989} to solve for single-phase wave motions of a perfect fluid. It has been applied extensively to the solution of finite amplitude wave propagation and wave breaking problems (see chapter 3 of \citet{Ma2010}). The perfect fluid assumption restricts WSIM to simulating breaking impact just prior to surface reconnection of the breaking crest tip to the forward face of the wave. However, its potential theory formulation enables it to simulate wave propagation in a CPU-efficient way, without the diffusion issues of viscous numerical codes. The simulation of wave generation and development of the onset of breaking events can be carried out with great precision~(e.g.~\cite{Fochesato2006,Fochesato2007}). WSIM has been validated extensively for wave evolution in deep and intermediate depth water and shows excellent energy conservation~\citep{Grilli1989,Grilli1990,Grilli1994,Grilli1996,Grilli1997,Grilli2001,Fochesato2006,Fochesato2007}.  

%

With reference to Table 1 of~\citep{Barthelemy2018}, we consider a suite of representative 2D and 3D gravity wave groups in deep and transitional waters. From the simulated wave profiles, we computed the evolution of the crest velocity $c$ as a function of the local crest steepness $S_{c}=k_c a_c$, where $a_c$ is the amplitude of the crest and $k_c=2\pi/L_c$ is the associated local wavenumber, with $L_c$ being the distance between the two zero-crossings adjacent to the crest. In each simulated case, the crest velocity $c$ has been normalised by the linear phase speed $c_{0}$.
For reference, we consider the celerity $c$ in~\eqref{eq:Stokes} of a $5^{th}$ order Stokes steady wave and the associated local steepness $S_c$ is computed from the theoretical profiles~\citep{Fenton1979,Fenton1985}. Three different methods (local, average and global) were used to obtain $c_0=\omega_0 / k_0$. The linear velocity $c_{0}$ is computed without assuming a dispersion relationship, using the spectral-peak frequency $\omega_{0}$ at the wavenumber~$k_{0}$. The local method values were found by computing $k_0$ from the spatial FFT at a arbitrary time and $\omega_0$ from a FFT time series centred on the position of the peak of the wave group at this given time. The average method determined $\omega_{0}$ from the ensemble mean FFT of the time series from 15 virtual height probes spaced uniformly between two crest (or trough) maxima. $k_{0}$ was determined from the ensemble mean FFT of $512$ surface profiles between two crest (or trough) maxima. The global method used a 2D FFT over the entire  space-time domain of the simulation to obtain the dispersion graph from which $\omega_{0}$ and $k_0$ were extracted. Refined values were found using an interpolating polynomial around the power spectral maxima. We observed that the values of $c_{0}$ obtained from the average and the global methods agree within $\pm 1\%$.

\begin{figure}[h]
\centering\includegraphics[scale=0.45]{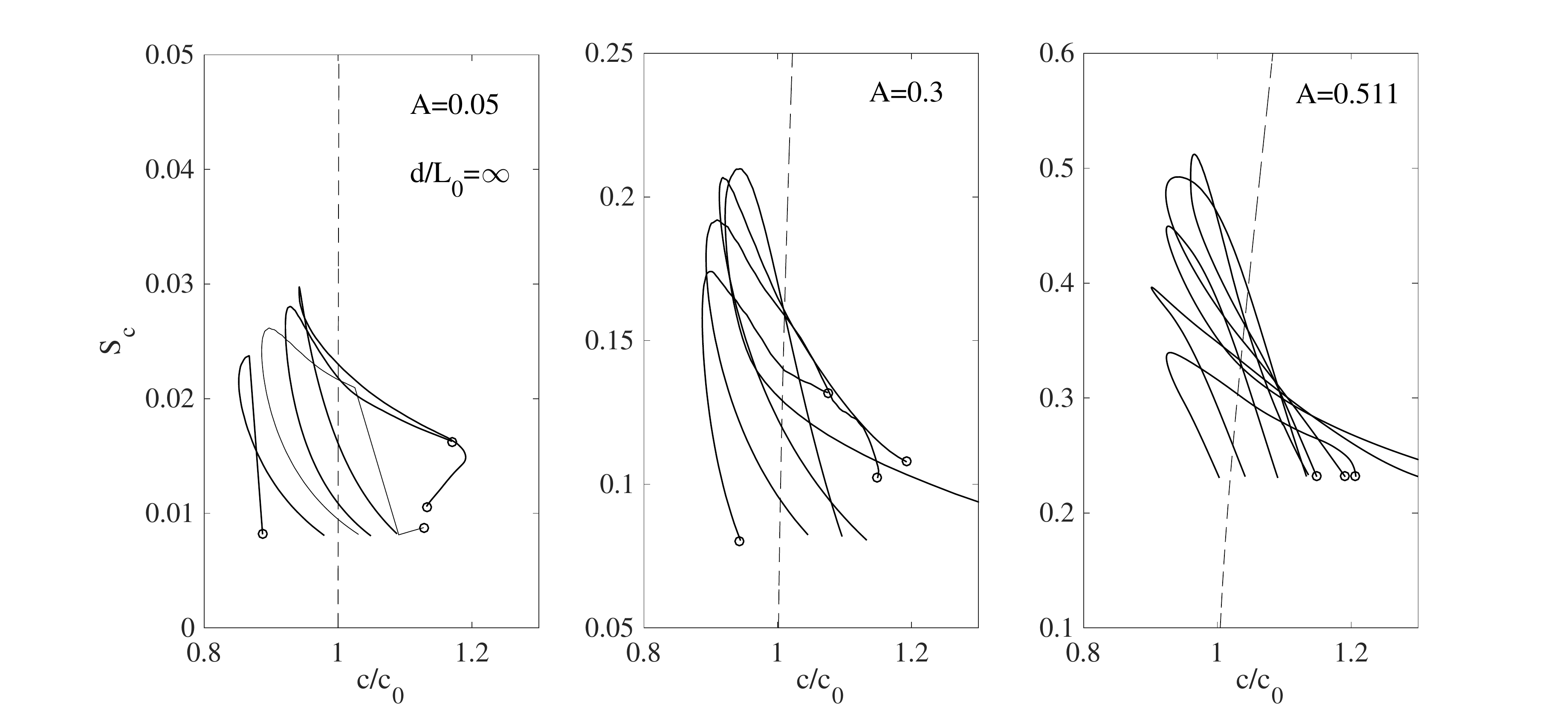}             
\caption{Hysteresis diagrams $S_c$~against~$c/c_0$ for 2D deep water C3N5 small steepness (left) and maximally steep nonbreaking (right) gravity wave groups. Here, $c_0$ is the reference linear deep-water phase speed of the dominant frequency~$\omega_0$. Dashed lines depict $5^{th}$~order Stokes prediction~\citep{Fenton1979,Fenton1985}.}
\label{FIGX}
\end{figure}

With reference to Table 1 of~\citep{Barthelemy2018}, consider the simulated 2D C3N5 wave packets in deep water. Hysteresis diagrams $S_c$~against~$c/c_0$ for C3N5 small (left) and moderate (center) steepness, as well as maximally steep non-breaking (right) gravity wave groups are shown in Figure~\ref{FIGX}. Here, $c_0$ is the reference linear deep-water phase speed of the dominant frequency~$\omega_0$. In particular, the left panel of this Figure depicts the same hysteresis diagram of a 2D C3N5 deep water gravity wave packet with small steepness~($A=0.05$).  The center panel of the same figure shows the hysteresis diagram for the same group with moderate steepness~($A=0.3$), whereas the maximum nonbreaking state~($A=0.511$) is shown in the left panel. Here, $A$ is the maximum local steepness $S_c$ observed in the wave group simulation. The prediction of crest velocity dependence on steepness based on $5^{th}$order Stokes theory~\citep{Fenton1979,Fenton1985} is shown by the dashed line. The crest velocities for the small steepness wave group have the largest range varying in the interval~$[0.85c_0-1.2c_0]$.  As the steepness increases, this dynamical range reduces especially for the highly nonlinear maximum nonbreaking case~(left panel), where the minimum crest velocities cluster more closely around $0.9c_0$. 

Two important findings for this generic class of wave packets are apparent. First, all the crests of the wave group experience a crest velocity slowdown. Second, in each wave group case there is a tendency for the minimum speed of the ensemble of crests to increase as the local steepness increases. The slowdown experienced by all the crests within a wave group is dependent on the steepness of the group and the leaning cycles have almost the same duration close to the dominant wave period, irrespective of the group length. This is expected as waves in a group are phase-locked.  In particular, for the small-steepness 2D wave group (see left panel of Figure~\ref{FIGX}) both small and large crests experience a comparable crest slowdown, indicating that the slowdown is practically insensitive to the degree of phase coherence/focus of each crest. In contrast, crests of the maximally nonbreaking wave group (see right panel of Figure~\ref{FIGX}) experience a slowdown that is smaller for steeper crests. In each panel, the hysteresis curves tend to bend upward as the maximum crest steepness increases, with a reduction in the associated crest speed slowdown and smaller dynamical ranges of crest speeds.

In summary, Figure~\ref{FIGX} shows that crests of deep-water gravity wave groups have a complex evolution, slowing down towards a similar minimum value. No major differences in the hysteresis curves between 2D and strongly converging 3D gravity wave groups are seen, leading to the important conclusion that crest slowdown behaviour is a generic feature of 2D and 3D unsteady gravity wave packets as discussed below. Even at lower steepness levels, similar patterns in crest slowdown are observed. Moreover, the slowdown of crests within steep gravity wave groups reduces as the local steepness increases, suggesting that nonlinear dispersion is a plausible mechanism for the slowdown reduction. In brief, crest slowdown reduces when waves steepen as they become less dispersive. 

\begin{figure}[h]
\centering\includegraphics[scale=0.45]{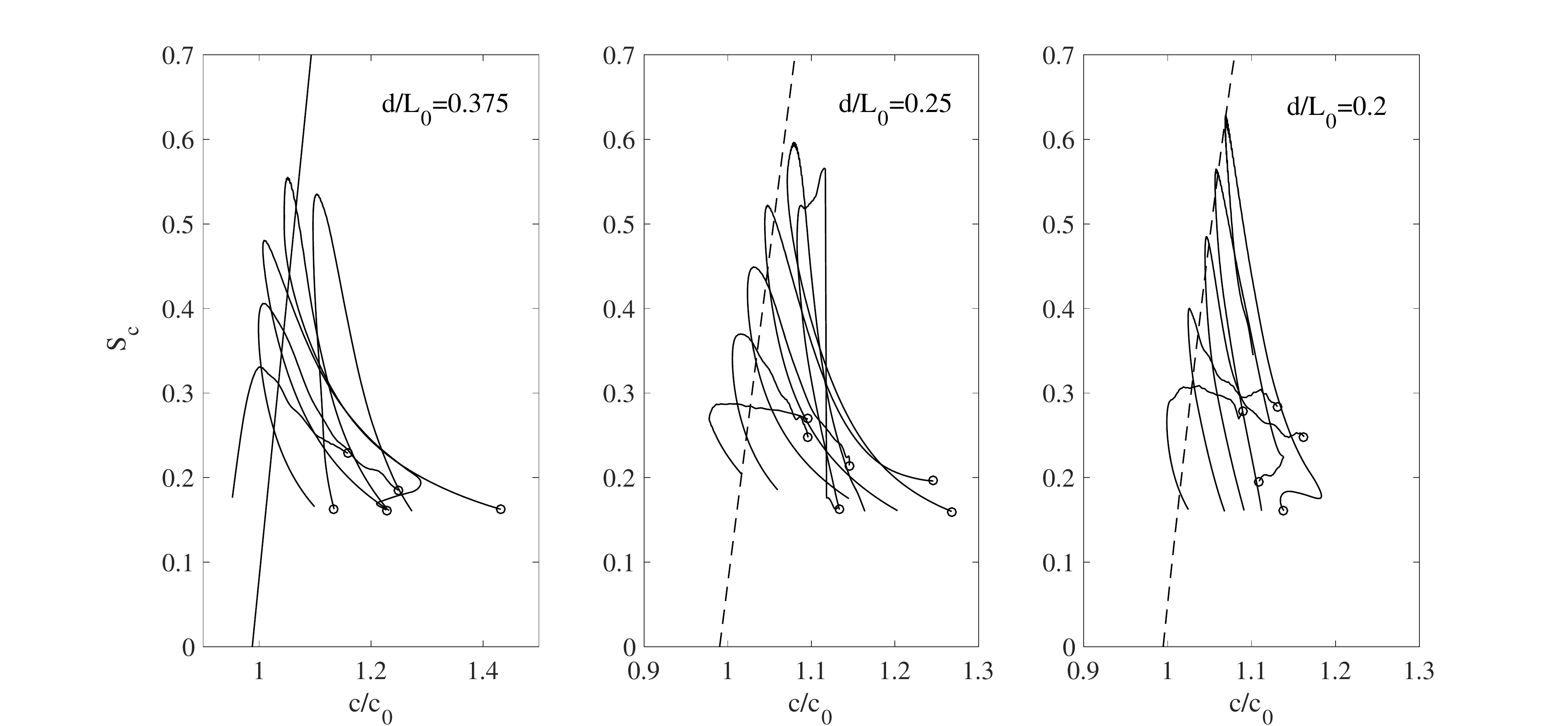} 
\caption{Hysteresis diagrams $S_c$~against~$c/c_0$ for 2D D3N9 gravity wave groups  from deep to transitional waters. Here, $c_0$ is the reference linear finite-depth phase speed of the dominant frequency~$\omega_0$. Dashed lines depict $5^{th}$~order Stokes prediction~\citep{Fenton1979,Fenton1985}.}
\label{FIGX1}
\end{figure}

Gravity waves also become less dispersive when they propagate over water of finite depth, and the crest slowdown is reduced. This is clearly seen in  Figure~\ref{FIGX1}, which shows the hysteresis diagrams $S_c$~vs.~$c/c_0$ of crests of a 2D D3N9 gravity wave packet from intermediate ($d/L_0=0.375$, left panel) to relatively shallow water depths ($d/L_0=0.25,0.2$, center and right panels).  Clearly, steeper crests experience a reduced dynamical range of values with smaller slowdown that decreases as the water depth reduces, that is, the hysteresis curves bend upward as water depth reduces. As a result, the associated crest speed slowdown reduces and  the crest speeds tend to align with the predicted $5^{th}$~order Stokes phase speed predictions~\citep{Fenton1979,Fenton1985} (dashed lines), especially as the maximum crest steepness increases. 

\begin{figure}[h]
\centering\includegraphics[scale=0.45]{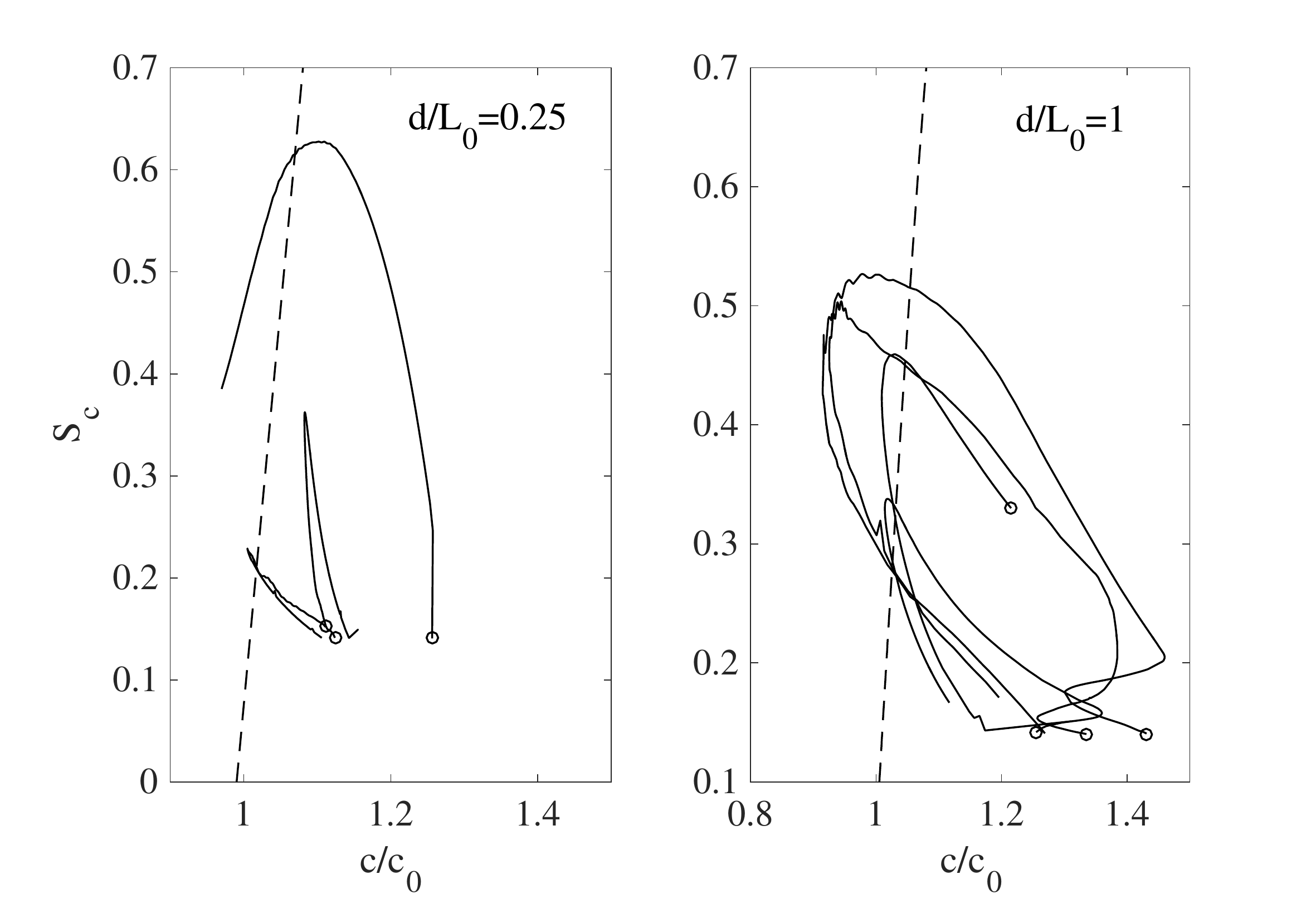} 
\caption{Hysteresis diagrams $S_c$~against~$c/c_0$ for 3D C3N9 gravity wave groups  in deep (right) and in transitional waters (left). Here, $c_0$ is the reference linear finite-depth phase speed of the dominant frequency~$\omega_0$. Dashed lines depict $5^{th}$~order Stokes prediction~\citep{Fenton1979,Fenton1985}.}
\label{FIGX2}
\end{figure}

Representative short-crested (3D) wave packets also exhibit similar crest slowdown hysteresis curve behavior, including the abatement of the slowdown effect when the waves travel in shallower water. This is seen in Figure~\ref{FIGX2}, which depicts resuls from model simulations of strongly laterally-convergent C3N9 (X10) wave packets with $d/L_0=1$ (left panel) and $d/L_0=0.25$ (right panel) respectively (see Table 2 of~\cite{Barthelemy2018}). 

These results support our hypothesis that reduced slowdown of steep crests is due to the nonlinear nature of gravity surface waves, which tend to be less dispersive when either their steepness increases or when they travel from deep to shallow water depths.

\section{Conclusions}

This study provides new insights into the geometric and kinematic properties of the generic crest slowdown of linear and highly nonlinear unsteady gravity wave groups. This phenomenon is essentially due to the dispersive nature of surface waves. Crest slowdown is observed routinely for linear (small steepness) gravity wave groups. It is enhanced as the spectral bandwidth increases, but it is essentially insensitive to the degree of phase coherence and focusing within a wave group. Further, the analysis of the kinematics of both simulated and observed ocean waves reveals a forward-to-backward leaning cycle for the individual wave crests within a wave group, leading to a systematic local crest slowdown of~$15$-$20\%$ of the reference wave velocity. Stereo ocean observations show that the crest slowdown reduces as the wave steepness increases.  This is consistent with our  hypothesis that the slowdown of steep crests is reduced because gravity surface waves tend to be less dispersive at higher steepness, or as they propagate over water of intermediate depth. This is supported by numerical simulations of unsteady evolution of 2D and 3D dispersive wave packets in both deep and intermediate waters. While our major focus is on gravity waves where generic crest slowdown occurs, results for capillary-dominated waves are also discussed, for which crest speeds cyclically speed up.
Overall, these findings contribute new insight into how water wave groups propagate when the constraint of steadiness is relaxed, providing the unsteady basic state for the onset of breaking or maximally-steep nonbreaking waves, in 2D and 3D gravity wave packets. In the context of wave breaking onset in deep and intermediate-depth water, we note that the observed significantly-reduced speed of initiation of breaker fronts reported from time to time in the literature~\citep{Banner2014} is explained by the crest slowdown properties described in this paper. This aspect has been investigated in detail in our companion paper which proposes a generic new threshold for predicting breaking onset~\citep{Barthelemy2018}.

\section{Acknowledgements}
WASS deployment at Acqua Alta was supported by CASE-EJ1P Joint Industry Project (N. 4545093).  The analysis of video measurements was also supported by ONR grant BAA 09-012: "Ocean Wave Dissipation and energy Balance (WAVE-DB): toward reliable spectra and first breaking statistics". MLB and XB were supported by the Australian Research Council under Discovery Projects DP0985602 and DP120101701. 


\section{Declaration of Interests}
The authors report no conflict of interest.

\section{Appendix A}
Consider $N$ Fourier modes at $k_j=j \Delta k$ separated by $\Delta k$ of order $O(1/N)$, then the wave surface displacements in Eq.~\eqref{etac} can be written as the Riemann sum 
\begin{equation}
\zeta\left(\alpha,N\right)=\sum_{j=1}^{N}a_{j}\cos(k_j x-\omega_j t+\alpha\theta_{j}),
\end{equation}
where $\omega_j=\omega(k_j)$ and the Fourier amplitude $a_{j}=S_{n}(k_j)\Delta k$ is of $O(1/N)$. The sum $\sum a_j$ of the N amplitudes is finite and of $O(1)$, and the sum $\sum a_j^2$  of the squared amplitudes is of $O(1/N)$. Note that~$\zeta$ in Eq.~\eqref{etac} differs from the classical definition of a Gaussian sea just by the definition of the Fourier amplitudes $a_j$. For Gaussian displacements, $a_j=\sqrt{2 S_n(k_j)\Delta k}$, and $a_j$ is of $O(1/\sqrt{N})$. 

Said that, 
\begin{equation}
\zeta\left(\alpha,N\right)=\sum_{j=1}^{N}a_{j}\left[ \cos(k_j x -\omega_j t)\beta_j - \sin( k_j x -\omega_j t )\gamma_j\right],
\end{equation}
where we have defined the random variables
\begin{equation}
\beta_j=\cos(\alpha \theta_j),\qquad \gamma_j=\sin(\alpha \theta_j). 
\end{equation}
The mean and variance of $\beta_j$ are
\begin{equation}
\mu_{\beta_j}=\frac{\sin(\pi\alpha)}{\pi\alpha}, \qquad \sigma^2_{\beta_j}=\frac{1}{2}+\frac{\sin(2\pi\alpha)}{4\pi\alpha}-\left(\frac{\sin(\pi\alpha)}{\pi\alpha}\right)^{2},
\end{equation}
and
\begin{equation}
\mu_{\gamma_j}=0, \qquad \sigma^2_{\gamma_j}=\frac{1}{2}-\frac{\sin(2\pi\alpha)}{4\pi\alpha},
\end{equation}
for any integers $j$. 
Then, the mean and variance of $\eta$ follow as 
\begin{equation}
\mu_{\zeta}(\alpha,N)=\sum_{j=1}^{N}a_{j} \frac{\sin(\pi\alpha)}{\pi\alpha}\cos(k_j x -\omega_j t),
\end{equation}
and
\begin{equation}
\sigma_{\zeta}^{2}(\alpha,N)=\frac{1}{2}\sum_{j=1}^{N}a_{j}^{2}\left[1+\cos(2 k_j x -2\omega_j t)\frac{\sin(2\pi\alpha)}{2\pi\alpha}-2\cos(k_j x -\omega_j t)^2\left(\frac{\sin(\pi\alpha)}{\pi\alpha}\right)^{2}\right]. 
\end{equation}
Note that $\int S_n(k)d k=1$. Then, $\sum_{j=1}^{N}a_{j}\rightarrow1$ in the limit of $N\rightarrow\infty$, since $a_j$ are of $O(1/N)$. Further, the sum of the squared amplitudes $\sum_{j=1}^{N}a_{j}^2\rightarrow0$ because the sum of $N$ terms of $O(1/N^2)$ is infinitesimally small as $N\rightarrow\infty$. 
In the same limit, the variance $\sigma_{\eta}^{2}$ vanishes and $\eta$ tends, with probability 1, to its expected value
\begin{equation}
\zeta=\lim_{N\rightarrow\infty}\mu_{\eta}(\alpha,N)=\sum_{j=1}^{N}a_{j}\frac{\sin(\pi\alpha)}{\pi\alpha}\cos(k_j x -\omega_j t),
\end{equation}
as a consequence of the strong law of large numbers~\citep{vanGroesen2014}. This has been confirmed by way of Monte Carlo simulations. In particular, we observed that convergence is practically attained when we use a number of Fourier modes $N>10^4$ to synthesize the wave group, and variability around the mean is observed for number of Fourier modes less than $N$. In integral form,
\begin{equation}
\zeta(x,t;\alpha)=\widetilde{h}(\alpha)\int S_n(k) \cos(k  x -\omega t) d k, 
\end{equation}
which is identical to the perfectly focusing group $\zeta$ in Eq.~\eqref{etac} except that now the crest amplitude is given by
\begin{equation}
\widetilde{h}(\alpha)=h\frac{\sin(\pi\alpha)}{\pi\alpha},
\end{equation}
and it is smaller than the maximal height $h=\widetilde{h}(\alpha=0)$ when the focus of the wave packet is perfect.

\bibliographystyle{jfm}
\bibliography{biblio-full}

\end{document}